\documentclass[12pt]{article}
\pdfoutput=1

\usepackage{amsmath}
\usepackage{amsfonts}
\usepackage{amscd}
\usepackage{amsthm}
\usepackage{setspace}

\usepackage{graphicx}
\usepackage{authblk}
\usepackage{caption}
\usepackage{ytableau}
\usepackage{mathtools}
\usepackage[all,cmtip]{xy}
\usepackage[numbers,compress]{natbib}

\usepackage[OT2,T1]{fontenc}
\DeclareSymbolFont{cyrletters}{OT2}{wncyr}{m}{n}
\DeclareMathSymbol{\Sha}{\mathalpha}{cyrletters}{"58}

\def\Z{\mathbb{Z}}

\def\C{\mathbb{C}}
\def\P{\mathbb{P}}

\def\til{\tilde}

\begin{document}

\begin{titlepage}

\begin{flushright}
YITP-16-98
\end{flushright}

\vskip 1cm

\begin{center}

{\large Discrete Gauge Groups in F-theory Models on \\ Genus-One Fibered Calabi-Yau 4-folds without Section}

\vskip 1.2cm

Yusuke Kimura$^1$
\vskip 0.4cm
{\it $^1$Yukawa Institute for Theoretical Physics, Kyoto University, Kyoto 606-8502, Japan}
\vskip 0.4cm
E-mail: kimura@yukawa.kyoto-u.ac.jp

\vskip 1.5cm
\abstract{\par We determine the discrete gauge symmetries that arise in F-theory compactifications on examples of genus-one fibered Calabi--Yau 4-folds without a section. We construct genus-one fibered Calabi-Yau 4-folds using Fano manifolds, cyclic 3-fold covers of Fano 4-folds, and Segre embeddings of products of projective spaces. Discrete $\Z_5$, $\Z_4$, $\Z_3$ and $\Z_2$ symmetries arise in these constructions. We introduce a general method to obtain multisections for several constructions of genus-one fibered Calabi-Yau manifolds. The pullbacks of hyperplane classes under certain projections represent multisections to these genus-one fibrations. We determine the degrees of these multisections by computing the intersection numbers with fiber classes. As a result, we deduce the discrete gauge symmetries that arise in F-theory compactifications. This method applies to various Calabi-Yau genus-one fibrations.}  

\end{center}
\end{titlepage}

\tableofcontents
\section{Introduction}
F-theory\cite{Vaf, MV1, MV2} is an extension of type IIB superstring theory, which provides a nonperturbative formulation. F-theory is compactified on genus-one fibered Calabi--Yau $m$-folds (with or without a section) where the axio-dilaton, $\tau=C_0+ie^{-\phi}$, is identified with the modular parameter of a genus-one fiber, enabling the axio-dilaton to have $SL_2(\Z)$ monodromy. 7-branes are magnetic sources for the RR 0-form $C_0$, and 7-branes are wrapped on the components of a discriminant locus, i.e., the codimension 1 locus in the base over which fibers degenerate.
\par Until recently, in most literature, elliptically fibered Calabi--Yau $m$-folds that admit a global section were used as compactification geometries for F-theory. (For models with a global section, see, for example, \cite{GW, MP, CGK, MPW, BGK, BMPWsection, CKP, BGKgauge, CGKP, BMPWSU(5), CKPaddendum, CKPS, BMPW, GK, LSW, CKPT, CGKPS, MP2}.) However, a Calabi--Yau $m$-fold with a torus fibration does not typically admit a section. Therefore, it is possible to consider genus-one fibered Calabi--Yau $m$-folds lacking a global section as compactification geometries for F-theory. In recent years, such F-theory compactifications on genus-one fibrations without a section have been considered in several studies, initiated in \cite{BM, MTsection}\footnote{F-theory models without a section were considered in \cite{BEFNQ, BDHKMMS}.}. For recent progress in F-theory compactifications on genus-one fibrations without a section, see also, for example, \cite{AGGK, KMOPR, GGK, MPTW, MPTW2, CDKPP, LMTW, GKK, K, K2, KCY4}. In \cite{MTsection}, Morrison and Taylor considered the Jacobian of a genus-one fibration without a section to show that the F-theory geometry without a section fits into the moduli of Weierstrass models. They observed that in the moduli of Weierstrass models, the geometry of genus-one fibrations includes the geometry of elliptic fibrations with a section. They argued that moving from an F-theory model with a global section to an F-theory model on genus-one fibrations without a section in the moduli space can be viewed as a Higgsing process, in which a $U(1)$ gauge symmetry is broken and a discrete gauge symmetry\footnote{For recent progress in discrete gauge symmetries, see, for example, \cite{KNPRR, ACKO, BS, HSsums, CIM, BISU, ISU, BCMRU, BCMU, MRV, HS, BRU, KKLM, BGKintfiber, HS2, GPR, CGP}.} remains. 
\par The Calabi--Yau genus-one fibrations with an identical Jacobian fibration, $J$, form a group referred to as the Tate--Shafarevich group, $\Sha(J)$. A genus-one fibration $Y$ and the Jacobian fibration $J(Y)$ have identical $\tau$ functions. Therefore, F-theory compactifications on distinct genus-one fibrations in $\Sha(J)$ describe physically equivalent theories. As discussed in \cite{MTsection}, the Tate--Shafarevich group $\Sha(J(Y))$ (of the Jacobian $J(Y)$) is identified with the discrete gauge group that forms in F-theory compactification on a Calabi--Yau genus-one fibration $Y$\cite{BDHKMMS}. Concretely, in F-theory compactified on a genus-one fibration with an $n$-section, a discrete $\Z_n$ symmetry arises as the remnant of a broken $U(1)^{n-1}$. For example, a discrete $\Z_2$ gauge symmetry arises in F-theory compactification on a genus-one fibration with a bisection, and when a genus-one fibration has a 3-section, a discrete $\Z_3$ symmetry arises in F-theory compactification.

\par In this note, we determine the discrete gauge symmetries that arise in F-theory compactifications on several constructions of genus-one fibered Calabi--Yau 4-folds without a section, by specifying the degrees of multisections. We build genus-one fibered Calabi-Yau manifolds that lack a global section using Fano manifolds. Fano manifolds are a generalization of (products of) projective spaces. We consider the following constructions of genus-one fibered Calabi-Yau 4-folds: \\
1. Intersection of two hyperplane classes in the product of two Fano 3-folds \\
2. Cyclic 3-fold covers of $\P^1$ times Fano 3-folds \\
3. Constructions involving Segre embeddings of $\P^2\times\P^2$ and $\P^1\times\P^1\times\P^1$\\
4. Hypersurfaces in $\P^2$ times Fano 3-folds \\
5. Complete intersections in $\P^3$ times Fano 3-folds \\
6. Double covers of $\P^1$ times Fano 3-folds
\par In the six constructions 1. $\sim$ 6., we particularly use Fano 3-folds of index 2, $V_d$ ($d=2,\cdots, 8$). These Fano 3-folds $V_d$ were discussed in \cite{Isk, Fujita1, Fujita2}. In this note, Fano 3-folds $V_d$ are referred to as {\it del Pezzo 3-folds of degree d}, following \cite{Fujita1, Fujita2}. We do not consider the del Pezzo 3-fold of degree 1, $V_1$, in this note. 
\par We determine the discrete gauge symmetries that form in F-theory compactifications on the above six constructions of genus-one fibered Calabi-Yau 4-folds. In this note, we introduce a general method which is applicable to all of the above six constructions of Calabi-Yau genus-one fibrations. 
\par As stated above, a discrete $\Z_n$ gauge symmetry forms in F-theory compactification on a Calabi-Yau genus-one fibration with a multisection of degree $n$. We introduce a general method to determine the degrees of multisections, as follows:
In each of the six Calabi-Yau constructions introduced above, genus-one fibers are embedded in a Fano manifold. The pullback of a hyperplane class in this Fano manifold to a Calabi-Yau space gives a multisection. The pullback of a point class in the base 3-fold, on the other hand, represents the fiber class. The degree of a multisection represents the number of intersection points with a fiber. (Therefore, a multisection of degree 1, which intersects with a fiber in a single point, is a global section.) Said differently, the intersection number of a multisection with the fiber class gives the degree of a multisection. Representing a multisection as the pullback of a hyperplane class enables to compute the intersection number with the fiber class. This method applies to all of the six Calabi-Yau constructions 1.$\sim$ 6. . Using this method, we deduce the discrete gauge groups that arise in F-theory compactified on the six constructions of Calabi--Yau 4-folds. 
\par This method is considerably general, and is also applicable to similar constructions of genus-one fibered Calabi-Yau manifolds. This method might be useful in determining discrete gauge symmetries in F-theory approach. 
\par The constructions 1.,2. and 3. provide novel examples of F-theory compactifications in which discrete symmetries form. In construction 1., discrete $\Z_2$, $\Z_3$, $\Z_4$ and $\Z_5$ gauge symmetries arise in F-theory compactifications employing del Pezzo 3-folds $V_2$, $V_3$, $V_4$ and $V_5$, respectively. In construction 2., a discrete $\Z_3$ symmetry arises in F-theory compactifications. In construction 3., a discrete $\Z_3$ gauge symmetry arises in F-theory compactifications involving Segre embedding of $\P^2\times\P^2$ into $\P^8$. A discrete $\Z_2$ symmetry arises in F-theory compactifications involving Segre embedding of $\P^1\times\P^1\times\P^1$ into $\P^7$. 
\par In construction 4., we consider hypersurfaces whose genus-one fibers are cubic hypersurfaces in $\P^2$. F-theory compactifications on Calabi-Yau genus-one fibrations whose fibers are cubic hypersurfaces in $\P^2$ were considered in \cite{KMOPR, CDKPP, K, KCY4}. In such F-theory compactifications, a discrete $\Z_3$ symmetry arises \cite{KMOPR, CDKPP}. 
\par In construction 5., we consider the complete intersections whose genus-one fibers are the complete intersections of two quadrics in $\P^3$. F-theory compactifications on Calabi-Yau genus-one fibrations, whose fibers are complete intersections of two quadric hypersurfaces in $\P^3$ were studied in \cite{BGKintfiber, K2}. In these F-theory compactifications, a discrete $\Z_4$ gauge symmetry arises \cite{BGKintfiber}. 
\par In construction 6., we consider Calabi-Yau spaces as double covers, whose genus-one fibers are double covers of $\P^1$ ramified over 4 points. F-theory compactifications of this kind were considered in \cite{MTsection, K2, KCY4}. In F-theory compactifications on these double covers, a discrete $\Z_2$ symmetry arises \cite{MTsection}. 
\par This paper is structured as follows: In Section \ref{sec 2}, we introduce the six constructions of genus-one fibered Calabi-Yau 4-folds that lack a global section. In Section \ref{sec 3}, we deduce discrete symmetries that form in F-theory compactifications on the six Calabi-Yau constructions. We represent multisections as pullbacks of hyperplane classes, and we compute the intersection numbers of these pullbacks of hyperplane classes with the fiber class to obtain the discrete symmetries. Constructions 1., 2. and 3. give novel examples of F-theory compactifications in which discrete symmetries form. In Section \ref{sec 4}, we state the concluding remarks. The continuous gauge theories and matter spectra in F-theory compactification on the special complete intersections whose K3 fibers are isomorphic to the Fermat quartic will be discussed in Appendix \ref{sec A}.

\section{Constructions of Genus-One Fibered Calabi-Yau 4-folds}
\label{sec 2}
In this section, we introduce constructions of genus-one fibrations without a section that are Calabi--Yau 4-folds. In Section \ref{sec 3}, we determine the discrete gauge symmetries that arise in F-theory compactifications on these genus-one fibrations. 

\subsection{Intersections of Two Hyperplane Classes in Product of Del Pezzo 3-folds}
\label{ssec 2.1}
In this section, we consider special Fano 3-folds, called the del Pezzo 3-folds of degree $d$, $V_d$, $d=2,\cdots, 8$, to construct genus-one fibered Calabi-Yau 4-folds. Structures of the del Pezzo 3-folds $V_d$, $d=2,\cdots, 8$ \cite{Isk, Fujita1, Fujita2} are displayed in Table \ref{Table del Pezzo} below. 

\begingroup
\renewcommand{\arraystretch}{1.5}
\begin{table}[htb]
\begin{center}
  \begin{tabular}{|c|c|} \hline
 Degree $d$ of del Pezzo 3-fold $V_d$ & Structure  \\ \hline
 2 & Double cover of $\P^3$  \\
 3 &  Cubic 3-fold in $\P^4$ \\ 
 4 &  Complete intersection of two quadrics in $\P^5$\\
 5 &  Intersection of three hyperplane sections in $Gr(2,5)\subset \P^9$\\ 
 6 &  $\P^1\times\P^1\times\P^1$ \\
 7 &  Blow-up of $\P^3$ at a point \\
 8 &  $\P^3$ \\ \hline   
\end{tabular}
\caption{\label{Table del Pezzo}List of del Pezzo 3-folds with degree $d$, $d=2, \cdots, 8$ \cite{Isk, Fujita1, Fujita2}. $Gr(2,5)$ denotes the complex Grassmannian of two-dimensional linear subspaces in $\C^5$. $Gr(2,5)$ is embedded inside $\P^9$ using the Pl\"ucker embedding; the restriction of a hyperplane section in $\P^9$ to $Gr(2,5)$ gives a hyperplane section of $Gr(2,5)$. The del Pezzo 3-fold of degree $8$ $V_8$ is the image of $\P^3$ embedded in $\P^9$ under the Veronese embedding.}
\end{center}
\end{table}  
\endgroup 

\par We consider the product of two del Pezzo 3-folds $V_D\times V_E$, and we consider the intersection of two hyperplane sections\footnote{Del Pezzo 3-fold $V_8=\P^3$ is seen as a subvariety embedded in $\P^9$ using the Veronese map of degree 2, $v_2$: $V_8=\P^3 \xhookrightarrow{} \P^9$. Hyperplane class $\mathcal{O}_{V_8}(1)$ in del Pezzo 3-fold $V_8$ is defined as the restriction of hyperplane class $\mathcal{O}_{\P^9}(1)$ in $\P^9$ to $V_8$: $\mathcal{O}_{V_8}(1):=\mathcal{O}_{\P^9}(1)|_{V_8}$. Therefore, $\mathcal{O}_{V_8}(1)=v_2^* \mathcal{O}_{\P^9}(1)=\mathcal{O}_{\P^3}(2)$.} $\mathcal{O}_{V_D\times V_E}(1,1)$ in the product $V_D\times V_E$. The resulting 4-fold $Y_4$ has the trivial canonical bundle; therefore it is a Calabi-Yau 4-fold. 
\par By construction, Calabi-Yau 4-fold $Y_4$ has projection $p$ onto $V_E$, and projection $q$ onto $V_D$. Fiber of projection $p$ is the intersection of two hyperplane classes $\mathcal{O}_{V_D}(1)$ in del Pezzo 3-fold $V_D$, therefore it is a genus-one curve. Similarly, fiber of projection $q$ is the intersection of two hyperplane classes $\mathcal{O}_{V_E}(1)$ in del Pezzo 3-fold $V_E$, i.e. a genus-one curve. Therefore, projections $p,q$ are genus-one fibrations. 
\par In this note, we focus on the case $D=2,3,4,5$ and $E=2, \cdots, 8$, and we choose $V_E$ to be the base 3-fold. Genus-one fiber is the intersection of two hyperplanes $\mathcal{O}_{V_D}(1)$ in del Pezzo 3-fold $V_D$, $D=2,3,4,5$. 
\par In Section \ref{ssec 3.1}, we deduce that the pullback of a hyperplane class in $V_D$
\begin{equation}
q^*\mathcal{O}_{V_D}(1)
\label{multi in 2.1}
\end{equation}
gives a multisection of degree $D$ to the fibration. Therefore, we find that a discrete $\Z_D$ gauge symmetry ($D=2,3,4,5$) forms in F-theory compactification on Calabi-Yau 4-fold $Y_4$. 
\par We note that $\P^1\times \P^2$ may be used as the base 3-fold, in place of $V_E$. For this case, Calabi-Yau 4-fold is built as the complete intersection of $\mathcal{O}_{V_D \times \P^1\times\P^2}(1,1,2)$ and $\mathcal{O}_{V_D \times \P^1\times\P^2}(1,1,1)$ hypersurfaces in $V_D \times \P^1\times\P^2$. Natural projection onto $\P^1\times \P^2$ gives a genus-one fibration, and the pullback (\ref{multi in 2.1}) gives a multisection of degree $D$, same as the case stated above, in which the base 3-fold is $V_E$. 
\par We show that a generic member of genus-one fibered Calabi-Yau 4-folds $Y_4$, constructed as complete intersections of two $\mathcal{O}_{V_D\times V_E}(1,1)$ classes in $V_D \times V_E$ ($D=2,3,\cdots,6,8$, $E=2,\cdots,8$), does not have a global section. (The base 3-fold of genus-one fibration is $V_E$.) To show that $Y_4$ lacks a global section, we prove that the Picard group $Pic(Y_4)$, which is isomorphic to the group of divisor classes in $Y_4$, is generated by the pullbacks of the divisors in $V_D$ and the pullbacks of the divisors in $V_E$. First, we assume that $D\ne 6,8$, $E\ne 6,7$. The cases where $D=6,8$ or $E=6,7$ are merely special cases, and similar arguments as that stated below apply to these special cases. Since 
\begin{equation}
H^1(V_D)=0, \hspace{5mm} H^1(V_E)=0,
\end{equation}
(these follow from the fact that $V_D$ and $V_E$ are Fano 3-folds), by K\"unneth formula, we find that 
\begin{equation}
H^2(V_D \times V_E, \Z)\cong \Z^2.
\end{equation}
By Lefschetz theorem, 
\begin{equation}
\label{rank Picard in 2.1}
H^2(Y_4, \Z)\cong H^2(V_D \times V_E, \Z)\cong \Z^2.
\end{equation}
Since $V_D$ and $V_E$ are Fano 3-folds, $H^1(V_D \times V_E, \Z)=0$, therefore by Lefschetz theorem $H^1(Y_4, \Z)=0$. This shows that $Pic^0(Y_4)=0$, and thus $Pic(Y_4)\cong NS(Y_4)$. By the Lefschetz (1,1) theorem $NS(Y_4)\cong H^2(Y_4, \Z)\cap H^{1,1}(Y_4)$, therefore the rank of the Picard group is bounded above by the rank of $H^2(Y_4, \Z)$, which is 2 by (\ref{rank Picard in 2.1}). The pullbacks of $\mathcal{O}(1)$ classes in $V_D$ and $V_E$ give divisors of $Y_4$, thus we conclude that these pullbacks generate the Picard group $Pic(Y_4)$. The pullback $p^*\mathcal{O}_{V_E}(1)$ is the pullback of a divisor in the base $V_E$; therefore, the pullback $p^*\mathcal{O}_{V_E}(1)$ is parallel to the fiber class $F$, namely the pullback $p^*\mathcal{O}_{V_E}(1)$ has an intersection number 0 with the fiber class $F$. On the other hand, as we will see in Section \ref{ssec 3.1}, the pullback $q^*\mathcal{O}_{V_D}(1)$ of $\mathcal{O}_{V_D}(1)$ class in $V_D$ gives a $D$-section, when $D=2,3,4,5$. Therefore, a divisor in $Y_4$, which is $nq^*\mathcal{O}_{V_D}(1)+mp^*\mathcal{O}_{V_E}(1)$ for some integers $n,m$, has an intersection number a multiple of $D$ with the fiber class $F$. ($(nq^*\mathcal{O}_{V_D}(1)+mp^*\mathcal{O}_{V_E}(1))\cdot F=nD$.) This shows that $Y_4$ has only multisections whose degrees are multiples of $D$, thus we conclude that $Y_4$ does not have a global section.
\par When $D\ne 6, 8$, $E=6$, the second Betti number of the del Pezzo 3-fold of degree 6 $V_6$ is $b_2(V_6)=3$, therefore we have
\begin{equation}
H^2(V_D\times V_E, \Z)\cong \Z^4.
\end{equation}
By Lefschetz theorem, we obtain
\begin{equation}
\label{V6 cohomology in 2.1}
H^2(Y_4, \Z)\cong \Z^4.
\end{equation}
For this case, the pullback $p^*\mathcal{O}_{V_6}(1)$ splits into three classes: $p^*\mathcal{O}_{\P^1\times\P^1\times\P^1}(1,0,0)$, $p^*\mathcal{O}_{\P^1\times\P^1\times\P^1}(0,1,0)$ and $p^*\mathcal{O}_{\P^1\times\P^1\times\P^1}(0,0,1)$. This comes from the fact that $V_6$ is the direct product of three projective lines. Therefore, $Y_4$ has four divisors: these three pullbacks, and the pullback $q^*\mathcal{O}_{V_D}(1)$. From (\ref{V6 cohomology in 2.1}), we find that the rank of the Picard group $Pic(Y_4)$ is bounded above by 4. Thus, we conclude that the four divisors that we found generate the Picard group $Pic(Y_4)$. Three pullbacks $p^*\mathcal{O}_{\P^1\times\P^1\times\P^1}(1,0,0)$, $p^*\mathcal{O}_{\P^1\times\P^1\times\P^1}(0,1,0)$ and $p^*\mathcal{O}_{\P^1\times\P^1\times\P^1}(0,0,1)$ are parallel to the fiber class $F$, and the pullback $q^*\mathcal{O}_{V_D}(1)$ has an intersection number $D$ with the fiber class $F$. Thus, every divisor in $Y_4$ has an intersection number a multiple of $D$ with the fiber class $F$. This shows that $Y_4$ does not have a global section. 
\par When $D\ne 6, 8$, $E=7$, the base 3-fold $V_7$ is blow-up of $\P^3$ at a point; therefore, the Picard group of $V_7$ $Pic(V_7)$ is generated by $\mathcal{O}_{V_7}(1)$ and the exceptional divisor $E$. Since $b_2(V_7)=2$, 
\begin{equation}
H^2(V_D \times V_7, \Z)\cong \Z^3,
\end{equation}
and by Lefschetz theorem 
\begin{equation}
H^2(Y_4, \Z)\cong \Z^3.
\end{equation}
Thus, the rank of the Picard group $Pic(Y_4)$ is bounded by 3. The pullbacks $p^*\mathcal{O}_{V_7}(1)$, $p^*E$ and $q^*\mathcal{O}_{V_D}(1)$ give divisors of $Y_4$, therefore we conclude that these pullbacks generate the Picard group $Pic(Y_4)$. $p^*\mathcal{O}_{V_7}(1)$ and $p^*E$ are pullbacks of divisors in the base 3-fold $V_7$, thus they are parallel to the fiber class $F$. It follows that any divisor in $Y_4$ has an intersection number a multiple of $D$ with the fiber class $F$. This shows that $Y_4$ does not have a section. 
\par When $D=6$, multisection $q^*\mathcal{O}_{V_6}(1)$ splits into three bisections. (We discuss this case from the perspective of Segre embeddings in Section \ref{ssec 2.3} and \ref{sssec 3.3.2}.) When $D=8$, multisection $q^*\mathcal{O}_{V_8}(1)$ represents a 4-section. Similar arguments as that stated above show that for these cases $Y_4$ does not have a global section. When $D=7$, $Y_4$ admits a global section.

\subsection{Cyclic 3-fold Covers of $\P^1$ times Del Pezzo 3-folds}
\label{ssec 2.2}
Cyclic 3-fold covers of $\P^1\times V_E$ ramified along a $\mathcal{O}_{\P^1\times V_E}(3,3)$ 3-fold are Calabi-Yau 4-folds. As in Section \ref{ssec 2.1}, $V_E$ denotes the del Pezzo 3-fold of degree $E$, with $E=2, \cdots, 8$. 
\par By construction, a cyclic 3-fold cover, $Y$, has projection $p$ onto $V_E$, and projection $q$ onto $\P^1$. Fiber of projection $p$ onto $V_E$ is a cyclic 3-fold cover of a rational curve $\P^1$ ramified over 3 points, which is a genus-one curve. Therefore, projection $p$ gives a genus-one fibration. 
\par In Section \ref{ssec 3.2}, we will find that the pullback of a hyperplane class in $\P^1$
\begin{equation}
q^*\mathcal{O}_{\P^1}(1)
\end{equation}
gives a 3-section to the fibration. Thus, a discrete $\Z_3$ gauge symmetry arises in F-theory compactification on cyclic 3-fold cover $Y$. 
\par We remark that the possible non-Abelian gauge symmetries on the 7-branes in F-theory compactifications on cyclic 3-fold covers of $\P^1\times V_E$ are highly constrained by the symmetry. Similar observations are made in \cite{K, KCY4}. Genus-one fiber has the automorphism group of order 3, therefore every smooth fiber is the Fermat curve. Indeed, a rational curve $\P^1$ with 3 fixed points has the constant moduli, therefore the complex structure of its cyclic 3-fold cover is unique. It follows that fibers have the j-invariant 0 throughout the base; as a result, singular fibers have the j-invariant 0. Thus, the possible fiber types\footnote{We use Kodaira's notation to denote fiber types. For the Kodaira-N\'eron classification of the singular fibers of elliptic surfaces, see \cite{Kod1, Kod2, Ner}.} are $II$, $IV$, $I^*_0$, $IV^*$ and $II^*$; the possible non-Abelian gauge symmetries that can arise on the 7-branes are: $SU(3)$, $SO(8)$, $E_6$ and $E_8$. 

\subsection{Constructions Involving Segre embeddings of $\P^2\times\P^2$ and $\P^1\times\P^1\times\P^1$}
\label{ssec 2.3}
We construct Calabi-Yau 4-folds using Segre embeddings of products of projective spaces. We consider the following two Segre embeddings: \\
\begin{eqnarray}
\P^2\times \P^2 & \xhookrightarrow{} & \P^8 \\ 
\P^1\times \P^1 \times \P^1 & \xhookrightarrow{} & \P^7 
\end{eqnarray}
We denote a quadric hypersurface in $\P^4$ by $Q_3$. We consider the direct product of $Q_3$, and $\P^2\times\P^2$ embedded in $\P^8$ under the Segre embedding: $Q_3\times\P^2\times\P^2$. This is a Fano 7-fold, and the intersection of three hyperplane sections $\mathcal{O}_{Q_3\times \P^2\times\P^2}(1,1,1)$ in $Q_3\times\P^2\times\P^2$ has the trivial canonical bundle; therefore, the intersection of three hyperplane classes $\mathcal{O}_{Q_3\times \P^2\times\P^2}(1,1,1)$ gives a Calabi-Yau 4-fold, $Y_{2,2}$. 
\par By construction, Calabi-Yau 4-fold $Y_{2,2}$ has projection $p_1$ onto $Q_3$. Fiber of this projection is the intersection of three hyperplane classes $\mathcal{O}_{\P^2\times \P^2}(1,1)$ in $\P^2\times\P^2$, which is a genus-one curve; therefore, projection $p_1$ is a genus-one fibration. Calabi-Yau 4-fold $Y_{2,2}$ also has projection $q_1$ onto $\P^2\times\P^2$. We will find in Section \ref{sssec 3.3.1} that the pullbacks of hyperplane classes $\mathcal{O}_{\P^2\times\P^2}(1,0)$ and $\mathcal{O}_{\P^2\times\P^2}(0,1)$ under projection $q_1$ are 3-sections. Thus, a discrete $\Z_3$ symmetry arises in F-theory compactification on Calabi-Yau 4-fold $Y_{2,2}$.
\par Next, we consider the Segre embedding of $V_6=\P^1\times \P^1 \times \P^1$ into $\P^7$. We consider the direct product of two del Pezzo 3-folds, $V_6$ and $V_E$ ($E=2, \cdots, 8$): $V_6\times V_E$. We take the intersection of two hyperplane classes $\mathcal{O}_{V_6\times V_E}(1,1)$ in the product $V_6\times V_E$; the canonical bundle of this intersection has the trivial bundle, thus this construction gives a Calabi-Yau 4-fold, $Y_{1,1,1}$. 
\par By construction, Calabi-Yau 4-fold $Y_{1,1,1}$ has natural projection $p_2$ onto $V_E$. Fiber of projection $p_2$ is the intersection of two hyperplane classes $\mathcal{O}_{V_6}(1)$ in $V_6$, which is a genus-one curve. Therefore, projection $p_2$ gives a genus-one fibration. Calabi-Yau 4-fold $Y_{1,1,1}$ also has projection $q_2$ onto $V_6=\P^1\times \P^1 \times \P^1$. We will see in Section \ref{sssec 3.3.2} that the pullbacks of hyperplane classes $\mathcal{O}_{\P^1\times \P^1 \times \P^1}(1,0,0)$, $\mathcal{O}_{\P^1\times \P^1 \times \P^1}(0,1,0)$ and $\mathcal{O}_{\P^1\times \P^1 \times \P^1}(0,0,1)$ under $q_2$ are bisections. Therefore, a discrete $\Z_2$ symmetry arises in F-theory compactification on $Y_{1,1,1}$. 
\par A similar proof as that given in Section \ref{ssec 2.1} shows that generic members of the constructed genus-one fibered Calabi-Yau 4-folds $Y_{2,2}$ and $Y_{1,1,1}$ do not have a global section.

\subsection{Hypersurfaces in $\P^2$ times Fano 3-folds}
\label{ssec 2.4}
We construct Calabi-Yau 4-folds by considering hypersurfaces in $\P^2$ times Fano 3-folds. For Fano 3-folds, we use the following spaces: del Pezzo 3-folds $V_E$ ($E=2, \cdots, 8$), and $\P^1\times \P^2$. 
\par Hypersurface class $\mathcal{O}_{\P^2\times V_E}(3,2)$ in the product $\P^2\times V_E$ is a Calabi-Yau 4-fold, $Y_4$. Calabi-Yau 4-fold $Y_4$ has natural projection $p$ onto $V_E$, and natural projection $q$ onto $\P^2$. Fiber of projection $p$ is a cubic hypersurface in $\P^2$, which is a genus-one curve. Therefore, projection $p$ is a genus-one fibration. The particular case $E=6$ is a (3,2,2,2) hypersurface in $\P^2\times\P^1\times\P^1\times\P^1$ \cite{KCY4}. 
\par Similarly, (3,2,3) hypersurfaces in $\P^2\times\P^1\times\P^2$ are genus-one fibered Calabi-Yau 4-folds. Base 3-fold is $\P^1\times\P^2$. 
\par We will find in Section \ref{ssec 3.4} that the pullback of hyperplane class $\mathcal{O}_{\P^2}(1)$ under projection $q$ gives a 3-section. Therefore, a discrete $\Z_3$ symmetry arises in F-theory compactifications on these hypersurface constructions of Calabi-Yau 4-folds. 
\par A similar proof as that given in Section \ref{ssec 2.1} shows that a generic member of the constructed genus-one fibered Calabi-Yau 4-folds $Y_4$ does not have a global section. 

\subsection{Complete Intersections in $\P^3$ times Fano 3-folds}
\label{ssec 2.5}
We build Calabi-Yau 4-folds as complete intersections in $\P^3$ times Fano 3-folds. Same as in Section \ref{ssec 2.4}, we use del Pezzo 3-folds $V_E$, $E=2, \cdots, 8$, and $\P^1\times\P^2$, for Fano 3-folds. 
\par Complete intersections of two $\mathcal{O}_{\P^3\times V_E}(2,1)$ classes in the product $\P^3\times V_E$ are Calabi-Yau 4-folds. Fiber of natural projection $p$ onto $V_E$ is a complete intersection of two quadrics in $\P^3$, which is a genus-one curve. Therefore, projection $p$ gives a genus-one fibration. The particular case $E=6$ is a (2,1,1,1) and (2,1,1,1) complete intersection in $\P^3\times\P^1\times\P^1\times\P^1$. 
\par A similar argument as that stated above shows that (2,1,2) and (2,1,1) complete intersections in $\P^3\times\P^1\times\P^2$ are genus-one fibered Calabi-Yau 4-folds. 
\par Complete intersection in $\P^3$ times Fano 3-folds has natural projection $q$ onto $\P^3$. We will find in Section \ref{ssec 3.5} that the pullback of hyperplane class $\mathcal{O}_{\P^3}(1)$ in $\P^3$ under $q$ gives a 4-section. Therefore, we find that a discrete $\Z_4$ gauge symmetry arises in F-theory compactifications on genetic members of complete intersections described above.  
\par A similar proof as that given in Section \ref{ssec 2.1} shows that a generic member of the constructed genus-one fibered Calabi-Yau complete intersections does not have a global section.
\par We remark that, for special members of (2,1,1,1) and (2,1,1,1) complete intersections in $\P^3\times\P^1\times\P^1\times\P^1$, and special members of (2,1,2) and (2,1,1) complete intersections in $\P^3\times\P^1\times\P^2$, given by:
\begin{eqnarray}
\label{eq special type in 2.5}
x^2_1+x^2_3+2t \, f \, x_2x_4 & = & 0 \\
x^2_2+x^2_4+2t \, g \, x_1x_3 & = & 0. \nonumber
\end{eqnarray}
the discrete symmetry becomes $\Z_2$. ($[x_1:x_2:x_3:x_4]$ represents homogeneous coordinates on $\P^3$. $t$ is the inhomogeneous coordinate on the first $\P^1$ in $\P^3\times \P^1\times \P^1 \times \P^1$, and is the inhomogeneous coordinate on $\P^1$ in $\P^3\times\P^1\times\P^2$. For (2,1,1,1) and (2,1,1,1) complete intersections in $\P^3\times\P^1\times\P^1\times\P^1$, $f,g$ are bidegree (1,1) polynomials on $\P^1\times\P^1$, where this $\P^1\times \P^1$ is the product of the last two $\P^1$'s in $\P^3\times \P^1\times \P^1 \times \P^1$. For (2,1,2) and (2,1,1) complete intersections in $\P^3\times\P^1\times\P^2$, $f,g$ are polynomials of degree 2 and degree 1 on $\P^2$, respectively. Simultaneous vanishing of the two equations in (\ref{eq special type in 2.5}) gives a complete intersection.) (2,1,1,1) and (2,1,1,1) complete intersections, and (2,1,2) and (2,1,1) complete intersections have K3 fibrations, with base surface being $\P^1\times\P^1$, and $\P^2$, respectively. K3 fibers of the complete intersections (\ref{eq special type in 2.5}) are isomorphic to the Fermat quartic surface, $\{x^4+y^4+z^4+w^4=0\} \subset \P^3$ \cite{K2}. As we will see in Section \ref{ssec 3.5}, for special complete intersections (\ref{eq special type in 2.5}), 4-sections split into pairs of bisections. As a result, the discrete symmetry that forms in F-theory compactifications becomes $\Z_2$. 
\par We prove that the special (2,1,1,1) and (2,1,1,1) complete intersections in $\P^3\times\P^1\times\P^1\times\P^1$, and special (2,1,2) and (2,1,1) complete intersections in $\P^3\times\P^1\times\P^2$, both given by (\ref{eq special type in 2.5}), do not have a rational section. The proof goes as follows: the natural projection of (2,1,1,1) and (2,1,1,1) complete intersection (\ref{eq special type in 2.5}) onto $\P^1\times\P^1$, where these $\P^1$'s are the last two $\P^1$'s in the product $\P^3\times\P^1\times\P^1\times\P^1$, and the natural projection of (2,1,2) and (2,1,1) complete intersection (\ref{eq special type in 2.5}) onto $\P^2$, have K3 fibers, described by the simultaneous vanishing of the following two equations in $\P^3\times\P^1$:
\begin{eqnarray}
\label{K3 fiber of special type in 2.5}
x^2_1+x^2_3+2t \,  x_2x_4 & = & 0 \\
x^2_2+x^2_4+2t \,  x_1x_3 & = & 0. \nonumber
\end{eqnarray}
K3 fiber (\ref{K3 fiber of special type in 2.5}) is isomorphic to the Fermat quartic surface. K3 fiber (\ref{K3 fiber of special type in 2.5}) is a (2,1) and (2,1) complete intersection in $\P^3\times\P^1$; therefore, it has natural projection onto $\P^1$, which gives a genus-one fibration of K3 surface (\ref{K3 fiber of special type in 2.5}). This genus-one fibration does not have a global section \cite{K2}. Therefore, K3 fibers (\ref{K3 fiber of special type in 2.5}) are genus-one fibered, but they lack a global section. If (2,1,1,1) and (2,1,1,1) complete intersection (\ref{eq special type in 2.5}), and (2,1,2) and (2,1,1) complete intersection (\ref{eq special type in 2.5}), have rational sections, these rational sections restrict to K3 fibers, giving global sections to K3 fibers (\ref{K3 fiber of special type in 2.5}). Therefore, the fact that K3 fibers (\ref{K3 fiber of special type in 2.5}) lack a global section implies that (2,1,1,1) and (2,1,1,1) complete intersection Calabi-Yau 4-folds (\ref{eq special type in 2.5}) and (2,1,2) and (2,1,1) complete intersection Calabi-Yau 4-folds (\ref{eq special type in 2.5}) do not admit a rational section. 

\subsection{Double Covers of $\P^1$ times Fano 3-folds}
\label{ssec 2.6}
The following double covers of the products $\P^1$ times Fano 3-folds are Calabi-Yau 4-folds: 
\begin{itemize}
\item double covers of $\P^1\times V_E$ ramified over a $\mathcal{O}_{\P^1\times V_E}(4,4)$ 3-fold 
\item double covers of $\P^1\times\P^1\times\P^2$ ramified over a (4,4,6) 3-fold  
\end{itemize}
Each of these double covers has projection $p$ onto the base 3-folds $B_3$, where the base 3-fold $B_3$ is the del Pezzo 3-fold of degree $E$, $V_E$ for the first double cover construction, and the base 3-fold $B_3$ is $\P^1\times\P^2$ for the second double cover construction, respectively. Fiber of projection $p$ is a double cover of $\P^1$ ramified over 4 point, which is a genus-one curve. Thus, projection $p$ gives a genus-one fibration. 
\par These double covers have natural projection $q$ onto $\P^1$. We will see in Section \ref{ssec 3.6} that the pullback of point class $\mathcal{O}_{\P^1}(1)$ in $\P^1$ under $q$ gives a bisection. Therefore, a discrete $\Z_2$ symmetry arises in F-theory compactifications on these double covers.

\section{Discrete Gauge Symmetry}
\label{sec 3}
In this section, we represent multisections of the Calabi--Yau genus-one fibrations as the pullbacks of hyperplane classes from the spaces in which genus-one fibers embed. By computing the intersection numbers of these pullbacks with fiber classes $F$, we determine the degrees of the multisections. Using this method, we deduce the discrete gauge symmetries that arise in F-theory compactifications on the six constructions of genus-one fibered Calabi--Yau 4-folds without a section that we introduced in Section \ref{sec 2}.  

\subsection{Discrete $\Z_2$, $\Z_3$, $\Z_4$ and $\Z_5$ symmetries on Intersections of Two Hyperplane Classes in Product of Del Pezzo 3-folds}
\label{ssec 3.1}
In Section \ref{ssec 2.1}, we constructed genus-one fibered Calabi-Yau 4-fold $Y_4$ as the intersection of two hyperplane sections $\mathcal{O}_{V_D\times V_E}(1,1)$ in the product of two del Pezzo 3-folds, $V_D\times V_E$, $D=2,3,4,5$, $E=2, \cdots, 8$. 
\par By construction, Calabi-Yau 4-fold $Y_4$ has natural projection $p$ onto $V_E$, and natural projection $q$ onto $V_D$. We use $V_E$ as the base 3-fold. Projection $p$ gives a genus-one fibration; genus-one fiber is the intersection of two hyperplane classes $\mathcal{O}_{V_D}(1)$ in $V_D$. The pullback of a point class in the base 3-fold $V_E$, $p^*\{{\rm pt}\}$, has self-intersection 0; therefore it represents the fiber class $F$. The pullback of a hyperplane class $\mathcal{O}_{V_D}(1)$ in $V_D$, $q^*\mathcal{O}_{V_D}(1)$, has an intersection number $D$ with fiber class $F= p^*\{{\rm pt}\}$, where $D$ is the degree of the del Pezzo 3-fold $V_D$. Thus, pullback
\begin{equation}
q^*\mathcal{O}_{V_D}(1)
\end{equation}
represents a $D$-section, $D=2,3,4,5$.
\par Therefore, we conclude that a discrete $\Z_5$ symmetry arises in F-theory compactifications on intersections of two hyperplane sections in $V_5\times V_E$. Discrete $\Z_4$, $\Z_3$ and $\Z_2$ symmetries arise in F-theory compactifications on intersections of two hyperplane sections in $V_4 \times V_E$, $V_3 \times V_E$ and $V_2 \times V_E$, respectively. 
\par Additionally, as stated in Section \ref{ssec 2.1}, when $D=8$ multisection $q^*\mathcal{O}_{V_8}(1)$ represents a 4-section. Thus, a discrete $\Z_4$ symmetry forms in F-theory compactifications on intersections of two hyperplane sections in $V_8\times V_E$.
\par As stated in Section \ref{ssec 2.1}, the base 3-fold $V_E$ may be replaced by $\P^1\times\P^2$. The results obtained above remain unchanged when the base 3-fold is replaced by $\P^1\times\P^2$.

\subsection{Discrete $\Z_3$ symmetry on Cyclic 3-fold Covers of $\P^1$ times Del Pezzo 3-folds}
\label{ssec 3.2}
Cyclic 3-fold covers of $\P^1\times V_E$ ramified over a $\mathcal{O}_{\P^1\times V_E}(3,3)$ 3-fold are Calabi-Yau 4-folds. $V_E$ denotes the del Pezzo 3-fold of degree $E$, with $E=2, \cdots, 8$. Such cyclic 3-fold covers have projection $p$ onto $V_E$, and projection $q$ onto $\P^1$. Projection $p$ gives a genus-one fibration; genus-one fiber is a cyclic 3-fold cover of a rational curve $\P^1$ ramified over 3 points. 
\par The pullback of a point class in the base 3-fold $V_E$, $p^*\{{\rm pt}\}$, has self-intersection 0; therefore it gives the fiber class $F$. The pullback of a point class $\mathcal{O}_{\P^1}(1)$ in $\P^1$, $q^*\mathcal{O}_{\P^1}(1)$, has an intersection number 3 with fiber class $F= p^*\{{\rm pt}\}$. Therefore, we conclude that the multisection
\begin{equation}
q^*\mathcal{O}_{\P^1}(1)
\end{equation}
is a 3-section. Thus, a discrete $\Z_3$ gauge symmetry forms in F-theory compactifications on cyclic 3-fold covers of $\P^1\times V_E$ ramified over a $\mathcal{O}_{\P^1\times V_E}(3,3)$ 3-fold. 

\subsection{Discrete $\Z_3$ and $\Z_2$ symmetries on Constructions Involving Segre Embeddings of $\P^2\times\P^2$ and $\P^1\times\P^1\times\P^1$}
\label{ssec 3.3}

\subsubsection{$\Z_3$ symmetry in constructions involving Segre embeddings of $\P^2\times\P^2$ into $\P^8$}
\label{sssec 3.3.1}
We consider the Segre embedding of $\P^2\times\P^2$ into $\P^8$: $\P^2\times\P^2\xhookrightarrow{} \P^8$. $Q_3$ is a quadric 3-fold in $\P^4$. We saw in Section \ref{ssec 2.3} that the intersection of three hyperplane sections $\mathcal{O}_{Q_3\times\P^2\times\P^2}(1,1,1)$ in the product $Q_3\times\P^2\times\P^2$ is a Calabi-Yau 4-fold, $Y_{2,2}$. $Y_{2,2}$ has projection $p_1$ onto $Q_3$, and projection $q_1$ onto $\P^2\times\P^2$. Projection $p_1$ gives a genus-one fibration; genus-one fiber is the intersection of three hyperplane classes $\mathcal{O}_{\P^2\times\P^2}(1,1)$ in $\P^2\times\P^2$. 
\par The pullback of a point class in the base 3-fold $Q_3$, $p_1^*\{{\rm pt}\}$, represents the fiber class $F$. The pullback of hyperplane class $\mathcal{O}_{\P^2\times\P^2}(1,0)$ in $\P^2\times\P^2$, $q^*_1\mathcal{O}_{\P^2\times\P^2}(1,0)$, has an intersection number 3 with fiber class $F$. Similarly, pullback $q^*_1\mathcal{O}_{\P^2\times\P^2}(0,1)$ has an intersection number 3 with fiber class $F$. Therefore, we find that pullbacks
\begin{equation}
q^*_1\mathcal{O}_{\P^2\times\P^2}(1,0), \hspace{5mm} q^*_1\mathcal{O}_{\P^2\times\P^2}(0,1)
\end{equation}
represent 3-sections. A discrete $\Z_3$ gauge group arises in F-theory compactification on $Y_{2,2}$. 

\subsubsection{$\Z_2$ symmetry in constructions involving Segre embeddings of $\P^1\times\P^1\times\P^1$ into $\P^7$}
\label{sssec 3.3.2}
We consider the Segre embedding of $V_6=\P^1\times\P^1\times\P^1$ into $\P^7$: $\P^1\times\P^1\times\P^1\xhookrightarrow{} \P^7$. Intersection of two hyperplane sections $\mathcal{O}_{V_6\times V_E}(1,1)$ in the product of two del Pezzo 3-folds, $V_6\times V_E$, $E=2, \cdots, 8$, is a Calabi-Yau 4-fold, $Y_{1,1,1}$. $Y_{1,1,1}$ has projection $p_2$ onto $V_E$, and projection $q_2$ onto $V_6$. Projection $p_2$ is a genus-one fibration, and genus-one fiber is the intersection of two hyperplanes $\mathcal{O}_{V_6}(1)$ in $V_6=\P^1\times\P^1\times\P^1$. 
\par The pullback of a point class in the base 3-fold $V_E$, $p_2^*\{{\rm pt}\}$, gives the fiber class $F$. The pullback of hyperplane class $\mathcal{O}_{\P^1\times\P^1\times\P^1}(1,0,0)$ in $V_6=\P^1\times\P^1\times\P^1$, $q^*_2\mathcal{O}_{\P^1\times\P^1\times\P^1}(1,0,0)$, has an intersection number 2 with fiber class $F$. Similarly, Pullbacks $q^*_2\mathcal{O}_{\P^1\times\P^1\times\P^1}(0,1,0)$ and $q^*_2\mathcal{O}_{\P^1\times\P^1\times\P^1}(0,0,1)$ have intersection number 2 with fiber class $F$. Therefore, pullbacks
\begin{equation}
q^*_2\mathcal{O}_{\P^1\times\P^1\times\P^1}(1,0,0), \hspace{2mm} q^*_2\mathcal{O}_{\P^1\times\P^1\times\P^1}(0,1,0), \hspace{2mm} q^*_2\mathcal{O}_{\P^1\times\P^1\times\P^1}(0,0,1)
\end{equation}
are bisections. Thus, a discrete $\Z_2$ gauge group forms in F-theory compactification on $Y_{1,1,1}$. 

\subsection{Discrete $\Z_3$ symmetry on Hypersurfaces in $\P^2$ times Fano 3-folds}
\label{ssec 3.4}
In Section \ref{ssec 2.4}, we considered the following two constructions of Calabi-Yau 4-fold as hypersurfaces in $\P^2$ times Fano 3-folds: 
\begin{itemize}
\item Hypersurfaces of class $\mathcal{O}_{\P^2\times V_E}(3,2)$ in $\P^2\times V_E$
\item (3,2,3) Hypersurfaces in $\P^2\times\P^1\times\P^2$
\end{itemize}
\par These Calabi-Yau hypersurfaces have projection $p$ onto $B_3$, and projection $q$ onto $\P^2$. For the first hypersurface, $B_3$ is the del Pezzo 3-fold $V_E$, and for the second hypersurface, $B_3$ is $\P^1\times\P^2$. Projection $p$ gives a genus-one fibration; a genus-one fiber is a cubic hypersurface in $\P^2$. The argument that follows to determine the degree of a multisection does not depend on the structure of $B_3$. 
\par The pullback of a point class in the base 3-fold $B_3$, $p^*\{{\rm pt}\}$, has self-intersection 0; therefore, it gives the fiber class $F$. The pullback of a line class $\mathcal{O}_{\P^2}(1)$ in $\P^2$, $q^*\mathcal{O}_{\P^2}(1)$, has an intersection number 3 with fiber class $F$. Therefore, pullback 
\begin{equation}
q^*\mathcal{O}_{\P^2}(1)
\end{equation}
represents a 3-section. A discrete $\Z_3$ symmetry arises in F-theory compactifications on these Calabi-Yau hypersurfaces. 

\subsection{Discrete $\Z_4$, $\Z_2$ symmetries on Complete Intersections in $\P^3$ times Fano 3-folds}
\label{ssec 3.5}
In Section \ref{ssec 2.5}, we considered the following complete intersections to construct Calabi-Yau 4-folds:
\begin{itemize}
\item Complete Intersections of two $\mathcal{O}_{\P^3\times V_E}(2,1)$ hypersurfaces in $\P^3\times V_E$
\item Complete Intersections of (2,1,2) and (2,1,1) hypersurfaces in $\P^3\times\P^1\times\P^2$.
\end{itemize}
\par By construction, these complete intersection Calabi-Yau 4-folds have projection $p$ onto $B^3$, and projection $q$ onto $\P^3$. $B_3$ is the del Pezzo 3-fold $V_E$, $E=2, \cdots, 8$, for the first complete intersection, and $B_3$ is $\P^1\times\P^2$ for the second complete intersection. Projection $p$ gives a genus-one fibration; a genus-one fiber is a complete intersection of two quadrics in $\P^3$. The argument that follows to determine the degree of a multisection does not depend on the structure of $B_3$.
\par The pullback of a point class in the base 3-fold $B_3$, $p^*\{{\rm pt}\}$, represents the fiber class $F$. The pullback of a surface class $\mathcal{O}_{\P^3}(1)$ in $\P^3$, $q^*\mathcal{O}_{\P^3}(1)$, has an intersection number 4 with fiber class $F$. Therefore, pullback
\begin{equation}
q^*\mathcal{O}_{\P^3}(1)
\end{equation}
represents a 4-section. A discrete $\Z_4$ gauge symmetry arises in F-theory compactifications on generic members of these complete intersections.
\par We remark that for special members of complete intersections, 4-sections split into pairs of bisections. We consider the special (2,1,1,1) and (2,1,1,1) complete intersections in $\P^3\times \P^1\times \P^1 \times \P^1$, and special (2,1,2) and (2,1,1) complete intersections in $\P^3\times\P^1\times\P^2$, given by 
\begin{eqnarray}
\label{eq special type in 3.5}
x^2_1+x^2_3+2t \, f \, x_2x_4 & = & 0 \\
x^2_2+x^2_4+2t \, g \, x_1x_3 & = & 0. \nonumber
\end{eqnarray}
\par We observe that for the special complete intersections (\ref{eq special type in 3.5}), 4-sections split into pairs of bisections. To observe this explicitly, we consider the following locus:
\begin{equation}
\label{locus in 3}
x_1=0.
\end{equation}
Along the locus (\ref{locus in 3}), the second equation in (\ref{eq special type in 3.5}) becomes
\begin{equation}
\label{special form of type 3 in 3}
x_2^2+x_4^2=0,
\end{equation}
which is independent of $t$ and $f,g$, i.e., independent of the coordinates on the base 3-fold $B_3$. Equation (\ref{special form of type 3 in 3}) splits into linear factors as follows:
\begin{equation}
x_2^2+x_4^2=(x_2+ix_4)(x_2-ix_4)=0,
\end{equation}
from which we find that $\{x_1=0, \hspace{1mm} x_2+ix_4=0\}$ and $\{x_1=0, \hspace{1mm} x_2-ix_4=0\}$ give bisections of a complete intersection (\ref{eq special type in 3.5}). This implies that a 4-section splits into a pair of two bisections, $\{x_1=0, \hspace{1mm} x_2+ix_4=0\}$ and $\{x_1=0, \hspace{1mm} x_2-ix_4=0\}$, for a complete intersection (\ref{eq special type in 3.5}). By applying a similar argument to three other loci, $x_2=0$, $x_3=0$, and $x_4=0$, we observe that a complete intersection (\ref{eq special type in 3.5}) has eight bisections. (We do not know if a complete intersection (\ref{eq special type in 3.5}) has more bisections.)
\par In summary, we have shown that the special complete intersections (\ref{eq special type in 3.5}) have bisections to the fibration, and from this result we deduce that a discrete $\Z_2$ gauge group arises in F-theory compactifications on the special complete intersections (\ref{eq special type in 3.5}).

\subsection{Discrete $\Z_2$ symmetry on Double Covers of $\P^1$ times Fano 3-folds}
\label{ssec 3.6}
In Section \ref{ssec 2.6}, we considered the following constructions of Calabi-Yau 4-folds as double covers of the products of $\P^1$ times Fano 3-folds:
\begin{itemize}
\item double covers of $\P^1\times V_E$ ramified over a $\mathcal{O}_{\P^1\times V_E}(4,4)$ 3-fold 
\item double covers of $\P^1\times\P^1\times\P^2$ ramified over a (4,4,6) 3-fold.  
\end{itemize}
\par By construction, these Calabi-Yau double covers have projection $p$ onto $B_3$, and projection $q$ onto $\P^1$. The base 3-fold $B_3$ is the del Pezzo 3-fold of degree $E$, $V_E$ for the first double cover construction, and $B_3$ is $\P^1\times\P^2$ for the second construction. The projection $p$ gives a genus-one fibration; genus-one fiber is a double cover of $\P^1$ ramified over 4 points. The argument that follows to determine the degree of a multisection does not depend on the structure of $B_3$. 
\par The pullback of a point class in the base 3-fold $B_3$, $p^*\{{\rm pt}\}$, has self-intersection 0; therefore, it represents the fiber class $F$. The pullback of a point class in $\P^1$, $q^*\mathcal{O}_{\P^1}(1)$, has an intersection number 2 with fiber class $F$. Therefore, pullback 
\begin{equation}
q^*\mathcal{O}_{\P^1}(1)
\end{equation}
is a bisection. A discrete $\Z_2$ symmetry arises in F-theory compactifications on these Calabi-Yau double covers.

\section{Conclusion}
\label{sec 4}
In this note, we investigated the discrete gauge symmetries that arise in F-theory compactifications on examples of genus-one fibered Calabi--Yau 4-folds without a section. We constructed genus-one fibered Calabi-Yau 4-folds that do not have a global section, by considering constructions using Fano manifolds, double covers, cyclic 3-fold covers and Segre embeddings of products of projective spaces. For these constructions, we obtained multisections as the pullbacks of hyperplanes classes in spaces in which fibers embed. We determined the degrees of these multisections by computing the intersection numbers with the fiber classes $F$, and we deduced the discrete gauge symmetries that form in F-theory compactifications on the six constructions of Calabi--Yau genus-one fibrations. 
\par In particular, the Calabi-Yau constructions that use the del Pezzo 3-folds (Section \ref{ssec 2.1}), cyclic 3-fold covers (Section \ref{ssec 2.2}), and the Segre embeddings of products of projective spaces (Section \ref{ssec 2.3}) provide novel examples of F-theory compactifications in which discrete gauge groups arise. Discrete $\Z_5$, $\Z_4$, $\Z_3$ and $\Z_2$ symmetries arise in the constructions that use the product of del Pezzo 3-folds, $V_D\times V_E$, depending on the degree of the del Pezzo 3-fold $V_D$. A discrete $\Z_3$ symmetry arises in F-theory compactifications on cyclic 3-fold covers of $\P^1\times V_E$ ramified over a $\mathcal{O}_{\P^1\times V_E}(3,3)$ 3-fold. A discrete $\Z_3$ symmetry arises in F-theory compactifications on the Calabi-Yau constructions that use the Segre embedding of $\P^2\times\P^2$ into $\P^8$. A discrete $\Z_2$ symmetry arises in F-theory compactifications on the Calabi-Yau constructions that use the Segre embedding of $\P^1\times\P^1\times\P^1$ into $\P^7$. 
\par For special complete intersection Calabi-Yau 4-folds in $\P^3\times\P^1\times\P^1\times\P^1$, and special complete intersection Calabi-Yau 4-folds in $\P^3\times\P^1\times\P^2$, whose K3 fibers are isomorphic to the Fermat quartic, we observed that 4-sections split into pairs of bisections. Consequently, the discrete symmetry that arises in F-theory compactifications becomes $\Z_2$ for these special cases. 
\par The method, introduced in Section \ref{sec 3} to determine the discrete gauge symmetries that arise in F-theory compactifications, is considerably general, and is applied to all six constructions of Calabi-Yau 4-folds. The same method applies to similar constructions of Calabi--Yau genus-one fibrations.

\section*{Acknowledgments}

We would like to thank Shigeru Mukai and Fumihiro Takayama for discussions. We are also grateful to the referee for improving this manuscript.

\newpage

\appendix
\section{Gauge Theories and Matter Spectra on (2,1,1,1) and (2,1,1,1) Complete Intersections in $\P^3\times\P^1\times\P^1\times\P^1$}
\label{sec A}
We determine the forms of the discriminant components, and the non-Abelian gauge groups arising on the 7-branes in F-theory compactifications on the special (2,1,1,1) and (2,1,1,1) complete intersections (\ref{eq special type in 2.5}) in $\P^3\times\P^1\times\P^1\times\P^1$. We derive the equations of the Jacobian fibrations and determine the Mordell--Weil group of the Jacobians. We find that the Mordell--Weil group has rank 0; therefore, F-theory models on the special (2,1,1,1) and (2,1,1,1) complete intersection Calabi--Yau 4-folds (\ref{eq special type in 2.5}) do not have a $U(1)$ gauge symmetry. The relationship between the gauge symmetry and the torsion part of the Mordell-Weil group is discussed in \cite{AM, MMTW}. By computing the Euler characteristic of the complete intersections (\ref{eq special type in 2.5}), we derive a condition imposed on a 4-form flux\footnote{\cite{GVW} studied 4-form flux. For recent studies of 4-form fluxes in F-theory, see, for example, \cite{BCV, MS, KMW, KMW2, CGK, BKL, CKPOR, SNW, LW}.} to cancel the tadpole. As discussed in \cite{KCY4}, to determine whether a consistent flux\cite{BB, SVW, W, DRS, LMTW} exists, we need to compute the self-intersections of intrinsic 2-cycles, which is technically considerably difficult. For this reason, in this study, we do not consider whether a consistent flux exists. We compute the potential matter spectra and potential Yukawa couplings. As it is undetermined whether a consistent flux choice exists, the contents of the spectra are potential matter and they {\it could} arise. Similar organization can be found in \cite{KCY4}.

\subsection{Forms of the Discriminant Components, Non-Abelian Gauge Groups on 7-Branes and Jacobian Fibration}
\label{ssec A.1}
In Section \ref{ssec 2.5} and \ref{ssec 3.5}, we discussed the special (2,1,1,1) and (2,1,1,1) complete intersections in $\P^3\times\P^1\times\P^1\times\P^1$, given by 
\begin{eqnarray}
\label{eq type 3 in A.1}
x^2_1+x^2_3+2t \, f \, x_2x_4 & = & 0 \\
x^2_2+x^2_4+2t \, g \, x_1x_3 & = & 0. \nonumber
\end{eqnarray}
The K3 fibers of the complete intersection Calabi--Yau 4-folds (\ref{eq type 3 in A.1}) are isomorphic to the Fermat quartic. Using a method similar to that in \cite{K2}, we derive the equation of the Jacobian fibrations of the complete intersections (\ref{eq type 3 in A.1}). 
\par We introduce a parameter $\lambda$, and subtract $\lambda$ times the second equation from the first equation in (\ref{eq type 3 in A.1}), as follows:
\begin{equation}
\label{eq built from type 3 in A.1}
x^2_1+x^2_3+2t \, f \, x_2x_4-\lambda(x^2_2+x^2_4+2t \, g \, x_1x_3).
\end{equation}
We arrange the coefficients of equation (\ref{eq built from type 3 in A.1}) into a symmetric 4 $\times$ 4 matrix, and compute the determinant of the matrix to obtain the equation of the Jacobian fibrations of the complete intersections (\ref{eq type 3 in A.1}) as follows: 
\begin{equation}
\tau^2=-t^2g^2\lambda^4+(t^4f^2g^2+1)\lambda^2-t^2f^2.
\label{jacobian ci}
\end{equation} 
The Jacobian fibrations (\ref{jacobian ci}) describe double covers of $\P^1\times\P^1\times\P^1\times\P^1$, where $\lambda$ and $t$ are the inhomogeneous coordinates on the first and the second $\P^1$'s, respectively, in $\P^1\times\P^1\times\P^1\times\P^1$, and $f,g$ are (1,1) polynomials on $\P^1\times\P^1$. (This $\P^1\times\P^1$ is the product of the third and the fourth $\P^1$'s in $\P^1\times\P^1\times\P^1\times\P^1$.) The Jacobians (\ref{jacobian ci}) are Calabi--Yau 4-folds. 
\par A complete intersection (\ref{eq type 3 in A.1}) and its Jacobian fibration (\ref{jacobian ci}) have the same discriminant loci and fiber types. Therefore, we can determine the forms of the discriminant components and the non-Abelian gauge groups in F-theory compactifications on the complete intersections (\ref{eq type 3 in A.1}) by studying the Jacobian fibrations (\ref{jacobian ci}). 
\par The discriminant of the Jacobian (\ref{jacobian ci}) is given by
\begin{equation}
\begin{split}
\Delta & =16f^2g^2t^4(f^2g^2t^4-1)^4 \\
       & =16f^2g^2t^4(fgt^2-1)^4(fgt^2+1)^4
\end{split}
\label{disc ci}
\end{equation} 
We can deduce the discriminant components from the vanishing of discriminant $\Delta$ (\ref{disc ci}) as follows: 
\begin{eqnarray}
\label{components in A.1}
E_1 & := & \{t=0\} \\ \nonumber
E_2 & := & \{t=\infty\} \\ \nonumber
D_1 & := & \{f=0\} \\ \nonumber
D_2 & := & \{g=0\} \\ \nonumber
D_3 & := & \{u=\infty\} \\ \nonumber
D_4 & := & \{v=\infty\} \\ \nonumber
D_5 & := & \{fgt^2=1\} \\ \nonumber
D_6 & := & \{fgt^2=-1\}. 
\end{eqnarray}
In (\ref{components in A.1}), we have used notations $u$ and $v$ to denote the inhomogeneous coordinates on the second and third $\P^1$'s, respectively, in the base 3-fold, $\P^1\times\P^1\times\P^1$. With these notations, $f,g$ are bidegree (1,1) polynomials in two variables, $u$ and $v$.  
\par We have 
\begin{equation}
E_i\cong \P^1\times\P^1 \hspace{5mm} (i=1,2)
\end{equation}
and
\begin{equation}
D_i\cong \P^1\times\P^1 \hspace{5mm} (i=3,4).
\end{equation}
$E_1$ and $E_2$ are parallel. $D_3\cap D_4$ is $\P^1$. We have 
\begin{equation}
E_i\cap D_j\cong \P^1 \hspace{5mm} (i=1,2,\hspace{2mm} j=3,4).
\end{equation}
A bidegree (1,1) curve in $\P^1\times\P^1$ is a rational curve $\Sigma_0\cong\P^1$. Therefore,
\begin{equation}
D_i\cong \P^1\times\Sigma_0\cong \P^1\times\P^1 \hspace{5mm} (i=1,2),
\end{equation} 
and 
\begin{equation}
\label{bulk rational}
E_i\cap D_j\cong \Sigma_0\cong \P^1 \hspace{5mm} (i,j=1,2).
\end{equation}
Two bidegree (1,1) curves in $\P^1\times\P^1$ intersect at 2 points; thus, $D_1\cap D_2$ is a disjoint sum of 2 $\P^1$'s. We omit components $D_{5,6}$. We display the forms of the irreducible components, $E_i$ ($i=1,2$) and $D_i$ ($i=1,2,3,4$), of the discriminant locus, and the forms of their intersections in Table \ref{table ci (3,1,1,1)}. 

\begingroup
\renewcommand{\arraystretch}{1.5}
\begin{table}[htb]
\begin{center}
  \begin{tabular}{|c|c|} \hline
Component & Topology \\ \hline
$E_i$ ($i=1,2$) & $\P^1\times\P^1$ \\
$D_i$ ($i=1,2,3,4$) & $\P^1\times\P^1$ \\ \hline
Intersections &  \\ \hline
$D_1\cap D_2$ & parallel 2 $\P^1$'s \\
$D_3\cap D_4$ & $\P^1$ \\
$E_i\cap D_j$ ($i=1,2$, $j=1,2,3,4$) & $\P^1$ \\ \hline   
\end{tabular}
\caption{Discriminant components of a complete intersection (\ref{eq type 3 in A.1}) of two (2,1,1,1) hypersurfaces in $\P^3\times \P^1\times \P^1 \times \P^1$, and their intersections. Components $D_5$ and $D_6$ are omitted.}
\label{table ci (3,1,1,1)}
\end{center}
\end{table}  
\endgroup    

\par We derive the non-Abelian gauge symmetries on the 7-branes. The Jacobian fibration (\ref{jacobian ci}) transforms into the extended Weierstrass form given by the equation
\begin{equation}
\label{extended Weierstrass in A.1}
y^2=\frac{1}{4}x^3-\frac{1}{2}(f^2g^2t^4+1)x^2+\frac{1}{4}(f^2g^2t^4-1)^2x.
\end{equation}
From the extended Weierstrass form, one can determine if a singular fiber is multiplicative (which corresponds to $I_n$ fibers) or additive (which corresponds to the other fiber types, i.e. $III$, $IV$, $II^*$, $III^*$, $IV^*$, $I^*_m$), by studying the coefficient of $x^2$ \footnote{See, for example, \cite{SchShio}.}. For example, when a singular fiber is at $t$, under some appropriate translation in $x$, an extended Weierstrass form transforms into another extended Weierstrass form $y^2=x^3+a_2x^2+a_4x+a_6$ in such a way that $t$ divides $a_4$ and $a_6$; when $t$ does not divide $a_2$, the fiber type at $t$ is multiplicative, and when $t$ divides $a_2$, the fiber type at $t$ is additive.
\par The discriminant $\Delta$ (\ref{disc ci}) vanishes under the conditions
\begin{equation}
\label{discriminant condition A.1}
fgt^2=0, \hspace{5mm} fgt^2-1=0, \hspace{5mm} fgt^2+1=0,
\end{equation}
and these conditions specify the locations of the singular fibers. The coefficient of $x^2$ in the extended Weierstrass form (\ref{extended Weierstrass in A.1})
\begin{equation}
-\frac{1}{2}(f^2g^2t^4+1)
\end{equation}
does not vanish under the conditions 
\begin{equation}
fgt^2-1=0, \hspace{5mm} fgt^2+1=0
\end{equation}
in (\ref{discriminant condition A.1}). Under the translation in $x$ that replaces $x$ with $x+1$, the extended Weierstrass form (\ref{extended Weierstrass in A.1}) transforms into 
\begin{equation}
\label{second extended Weierstrass in A.1}
y^2=\frac{1}{4}x^3+(\frac{1}{4}-\frac{1}{2}f^2g^2t^4)x^2+(\frac{1}{4}f^4g^4t^8-\frac{3}{2}f^2g^2t^4)x+\frac{1}{4}f^4g^4t^8-f^2g^2t^4.
\end{equation}
The coefficient of $x^2$ in the extended Weierstrass form (\ref{second extended Weierstrass in A.1})
\begin{equation}
\frac{1}{4}-\frac{1}{2}f^2g^2t^4
\end{equation}
does not vanish under the condition
\begin{equation}
fgt^2=0
\end{equation}
in (\ref{discriminant condition A.1}). Thus, we conclude that singular fibers are multiplicative. Therefore, the fiber types are $I_n$ for some $n$ \footnote{One can also see this fact by completing the cube to transform the extended Weierstrass form (\ref{extended Weierstrass in A.1}) into the Weierstrass form, and by studying the coefficients of the obtained Weierstrass form.}. The fiber types can be determined from the multiplicities of the zeros of discriminant $\Delta$ (\ref{disc ci}). The fiber type on components $E_1$ and $E_2$ is $I_4$. We consider the translation in $y$, $\til{y}=y+(t+\frac{1}{2})x+fgt^2$, for extended Weierstrass form (\ref{second extended Weierstrass in A.1}) to see that $I_4$ fibers on components $E_1$ and $E_2$ are of split type. The corresponding gauge groups on the 7-branes wrapped on components $E_1$ and $E_2$ are $SU(4)$. The fiber type on components $D_1$ and $D_2$ is $I_2$. Similarly, the fiber type on components $D_3$ and $D_4$ is $I_2$. The corresponding gauge groups on the 7-branes wrapped on $D_1$, $D_2$, $D_3$, and $D_4$ are $SU(2)$. The fiber type on components $D_5$ and $D_6$ is $I_4$. We consider the translation in $y$, $\til{y}=y-x$, for extended Weierstrass form (\ref{extended Weierstrass in A.1}) to find that $I_4$ fibers on components $D_5$ and $D_6$ are of split type. The corresponding gauge groups on the 7-branes wrapped on $D_5$ and $D_6$ are $SU(4)$. We show the results in Table \ref{table ci gauge}. 

\begingroup
\renewcommand{\arraystretch}{1.5}
\begin{table}[htb]
\begin{center}
  \begin{tabular}{|c|c|c|} \hline
Component & Fiber type & non-Abel. Gauge Group \\ \hline
$E_{1,2}$ & $I_4$ & $SU(4)$ \\ \hline
$D_{1,2,3,4}$ & $I_2$ & $SU(2)$ \\ \hline 
$D_{5,6}$ & $I_4$ & $SU(4)$ \\ \hline  
\end{tabular}
\caption{Singular fiber types and the corresponding gauge groups on the discriminant components of a complete intersection (\ref{eq type 3 in A.1}).} 
\label{table ci gauge}
\end{center}
\end{table}  
\endgroup

\subsection{Mordell--Weil Group of the Jacobian Fibration}
In \ref{ssec A.1}, we deduced that the Jacobian fibration of a complete intersection Calabi--Yau 4-fold (\ref{eq type 3 in A.1}) is given by the equation 
\begin{equation}
\tau^2=-t^2g^2\lambda^4+(t^4f^2g^2+1)\lambda^2-t^2f^2.
\label{jacobian ci in A.2}
\end{equation}
We compute the Mordell--Weil group of the Jacobian fibration (\ref{jacobian ci in A.2}). 
\par The projection from the Jacobian $(\ref{jacobian ci in A.2})$ onto $\P^1 \times \P^1$ (on which polynomials $f,g$ are defined) gives a K3 fibration, whose fiber is given by the following equation:
\begin{equation}
\tau^2=-t^2\lambda^4+(t^4+1)\lambda^2-t^2.
\label{jacobian K3 in A.2}
\end{equation}
The Mordell--Weil group of the K3 surface (\ref{jacobian K3 in A.2}) was determined to be $\Z_4\times\Z_4$ in \cite{K2}. We consider a specialization from the Jacobian fibration (\ref{jacobian ci in A.2}) to its K3 fiber (\ref{jacobian K3 in A.2}), which is equivalent to selecting a point in the base surface, $\P^1\times\P^1$, to deduce that the Mordell--Weil group of the Jacobian fibration (\ref{jacobian ci in A.2}) is isomorphic to that of the K3 surface (\ref{jacobian K3 in A.2}). Therefore, we conclude that the Mordell--Weil group of the Jacobian fibration (\ref{jacobian ci in A.2}) is $\Z_4\times\Z_4$. Thus, the global structure of the non-Abelian gauge group is 
\begin{equation}
SU(4)^4\times SU(2)^4 / \Z_4\times \Z_4.
\end{equation}
\par In particular, the Mordell--Weil group of the Jacobian (\ref{jacobian ci in A.2}) has rank 0. Thus, F-theory compactifications on the complete intersection Calabi--Yau 4-folds (\ref{eq type 3 in A.1}) do not have a $U(1)$ gauge field.  

\subsection{Euler Characteristic and Condition on 4-Form Flux to Cancel the Tadpole}
As stated earlier, we do not discuss whether a consistent flux exists. By computing the Euler characteristic of (2,1,1,1) and (2,1,1,1) complete intersection Calabi--Yau 4-folds in $\P^3\times\P^1\times\P^1\times\P^1$, we derive a condition imposed on the self-intersection of a 4-form flux $G_4$ to cancel the tadpole.
\par We compute the Euler characteristic $\chi(Y)$ of a (2,1,1,1) and (2,1,1,1) complete intersection Calabi--Yau 4-fold $Y$. An exact sequence, 
\begin{equation}
\label{SES}
\begin{CD} 
0 @>>> \mathcal{T}_Y @>>> \mathcal{T}_{\P^3\times \P^1\times\P^1\times\P^1}|_Y @>>> \mathcal{N}_Y @>>> 0 
\end{CD}
\end{equation}
gives the following equality: 
\begin{equation}
\label{ci Chern}
c(\mathcal{T}_Y)=\frac{c(\mathcal{T}_{\P^3\times \P^1\times\P^1\times\P^1})|_Y}{c(\mathcal{N}_Y)}.
\end{equation}
In the sequence (\ref{SES}), $\mathcal{T}_Y$ is the tangent bundle of the (2,1,1,1) and (2,1,1,1) complete intersection Calabi--Yau 4-fold $Y$, and $|_Y$ represents the restriction to $Y$. $\mathcal{N}_Y$ is the normal bundle resulting from natural embedding of tangent bundle $\mathcal{T}_Y$ into tangent bundle $\mathcal{T}_{\P^3\times \P^1\times \P^1 \times \P^1}$ of the ambient space, $\P^3\times \P^1\times \P^1 \times \P^1$.
\begin{equation}
\mathcal{N}_Y\cong \mathcal{O}(2,1,1,1)^{\oplus2},
\end{equation}
thus,
\begin{equation}
c(\mathcal{N}_Y)=(1+2x+y+z+w)^2.
\end{equation}
We have
\begin{equation}
c(\mathcal{T}_{\P^3\times \P^1\times\P^1\times\P^1})|_Y=(1+4x+6x^2+4x^3)(1+2y)(1+2z)(1+2w)|_Y.
\end{equation}
From equation (\ref{ci Chern}), we find that the Euler characteristic $\chi(Y)$, which is equal to the top Chern class of $c(Y)$, is
\begin{equation}
\chi(Y)=864.
\end{equation}
This is divisible by 24, and we have 
\begin{equation}
\frac{\chi(Y)}{24}=\frac{864}{24}=36.
\end{equation}  
\par Additionally, we find the second Chern class $c_2(Y)$ from equation (\ref{ci Chern}) as follows:
\begin{equation}
c_2(Y)=(2x^2+4xy+4xz+4xw+2yz+2yw+2zw)|_Y.
\end{equation}
This is even; thus, the quantization condition for a 4-form flux $G_4$\cite{W} reduces to  
\begin{equation}
G_4\in H^4(Y,\Z).
\end{equation} 
A bound on the self-intersection of a 4-form flux $G_4$ to cancel the tadpole is
\begin{equation}
\label{bound 4form in A.3}
N_3=\frac{\chi(Y)}{24}-\frac{1}{2}G_4\cdot G_4=36-\frac{1}{2}G_4\cdot G_4\ge 0
\end{equation} 
with $N_3$ the number of 3-branes minus anti 3-branes.

\begingroup
\renewcommand{\arraystretch}{1.5}
\begin{table}[htb]
\begin{center}
  \begin{tabular}{|c|c|c|} \hline
CY 4-fold $Y$ & Euler char. $\chi(Y)$ & $\frac{\chi(Y)}{24}$ \\ \hline
(2,1,1,1) and (2,1,1,1) complete intersection & 864 & 36 \\ \hline  
\end{tabular}
\caption{Euler characteristic of complete intersections of (2,1,1,1) and (2,1,1,1) hypersurfaces in $\P^3\times\P^1\times\P^1\times\P^1$.}
\label{table Euler char ci}
\end{center}
\end{table}  
\endgroup   
 
\subsection{Potential Matter Fields and Yukawa Couplings}
We deduce the potential matter fields on 7-branes and along matter curves, and Yukawa couplings. As stated earlier, it is undetermined whether a consistent flux exists. Therefore, we can only say that the matter spectra (and Yukawa couplings) we compute {\it could} arise.  
\par A deformation of the singularity associated with a gauge group $G$ generates matter fields on 7-branes\cite{BIKMSV, KV}. As stated in \cite{BHV1}, this corresponds to the breaking of the gauge group $G$ on 7-branes to a subgroup $\Gamma$, with maximal inclusion 
\begin{equation}
\Gamma\times H\subset G.
\end{equation}
We focus on the case in which $H$ is $U(1)$. 
\par We focus on the bulk component, $E_1$, of the discriminant locus. (Component $E_1$ is discussed in \ref{ssec A.1}.) We abbreviate $E_1$ to $E$. Let $\mathcal{L}$ be a supersymmetric line bundle on the bulk $E$. As $E\cong\P^1\times\P^1$ (see Table \ref{table ci (3,1,1,1)}), $\mathcal{L}\cong\mathcal{O}(a,b)$ for some integers $a,b\in\Z$. As argued in \cite{BHV1}, integers $a,b$ are required to satisfy the following inequality:
\begin{equation}
ab<0
\end{equation} for the line bundle $\mathcal{L}\cong\mathcal{O}(a,b)$ to be supersymmetric. 
\par Suppose $\Gamma$ has a representation $\tau$ with $U(1)$ charge $n$. Then, the line bundle associated to a matter in the representation $\tau$ is ${\mathcal{L}}^n$. The generation of matters in the representation $\tau$ of $\Gamma$ on the bulk $E$ is given by the following equation\cite{BHV1}: 
\begin{equation}
\label{generation eqn}
n_{\tau}-n_{\tau^*}=-\int_E c_1(E)c_1(\mathcal{L}^n)=-n\int_E c_1(E)c_1(\mathcal{L}).
\end{equation}
\par When the gauge group $SU(4)$ on the bulk $E$ breaks to $SU(3)$ under 
\begin{equation}
SU(4)\supset SU(3)\times U(1),
\end{equation}    
adjoint {\bf 15} of $SU(4)$ decomposes as\cite{Sla} 
\begin{equation}
{\bf 15}={\bf 8}_0+{\bf 3}_{-4}+\overline{\bf 3}_4+{\bf 1}_0.
\end{equation}
Thus, chiral matters {\bf 3} (could) arise on the bulk $E$, and their generation is given by
\begin{equation}
\begin{split}
n_{\bf 3}-n_{\overline{\bf 3}} & =-\int_Ec_1(E)c_1(\mathcal{L}^{-4})=4(2x+2y)(ax+by) \\
 & =8(a+b).
\end{split}
\end{equation}
\par We compute the generation of matter fields localized along the matter curve $E\cap D_i$, $i=1,2$. We saw in \ref{ssec A.1} that the matter curve $E\cap D_i$,  $i=1,2$, is isomorphic to a rational curve $\Sigma_0$, $E\cap D_i\cong\Sigma_0$, $i=1,2$. (See equation (\ref{bulk rational}) in \ref{ssec A.1}.) $\Sigma_0$ denotes a Riemann surface of genus 0, namely, a rational curve $\P^1$. 
\par $E\cap D_i$, $i=1,2$, is a bidegree (1,1) curve in $\P^1\times\P^1$. Therefore, the restriction $\mathcal{L}_{\Sigma_0}$, of the line bundle $\mathcal{L}\cong \mathcal{O}(a,b)$ to the matter curve $E\cap D_i=\Sigma_0$, $i=1,2$, is 
\begin{equation}
\mathcal{L}_{\Sigma_0}\cong \mathcal{O}_{\Sigma_0}(a+b).
\end{equation}
Matter field {\bf 6} localized along the matter curve $E\cap D_i\cong\Sigma_0$, $i=1,2$, decomposes as 
\begin{equation}
{\bf 6}={\bf 3}_{2}+\overline{\bf 3}_{-2}.
\end{equation}
We apply the Riemann--Roch theorem to obtain 
\begin{equation}
\begin{split}
n_{\bf 3} & =h^0(K^{1/2}_{\Sigma_0}\otimes \mathcal{L}^{2}_{\Sigma_0}) \\
           & =h^0(\mathcal{O}_{\Sigma_0}(2(a+b)-1)) \\
           & =\left\{
   \begin{aligned}
       2(a+b) \hspace{5mm} (a+b\ge0)\\
       0 \hspace{5mm} (a+b<0)
   \end{aligned}
   \right.
\end{split}
\end{equation}
Similarly,
\begin{equation}
\begin{split}
n_{\overline{\bf 3}} & =h^0(K^{1/2}_{\Sigma_0}\otimes \mathcal{L}^{-2}_{\Sigma_0}) \\
           & =h^0(\mathcal{O}_{\Sigma_0}(-2(a+b)-1)) \\
           & =\left\{
   \begin{aligned}
       -2(a+b) \hspace{5mm} (a+b\le0)\\
       0 \hspace{5mm} (a+b>0)
   \end{aligned}
   \right.
\end{split}
\end{equation}
Thus, when $a+b>0$, {\bf 3} (could) localize along the matter curve $E\cap D_i\cong\Sigma_0$, $i=1,2$. When $a+b<0$, matter fields $\overline{\bf 3}$ (could) localize along the matter curve $E\cap D_i\cong\Sigma_0$, $i=1,2$.
\par As discussed in \cite{BHV1}, Yukawa coupling arises from the interactions of the following three cases: i)Two matter fields on a matter curve and a matter on a bulk, ii)Three fields on a bulk, and iii)Matters along three matter curves intersecting at one point. $E\cong\P^1\times\P^1$ is a Hirzebruch surface; thus, Yukawa coupling does not arise from the interaction of the second case on the bulk $E$, as stated in \cite{BHV1}. We focus on Yukawa couplings arising from the first case. 
\par When $a+b>0$, matter field {\bf 3} on the bulk $E$ and two matter fields {\bf 3} localized along the matter curve $\Sigma_0$ generate the following Yukawa coupling: 
\begin{equation}
\label{Yukawa 1 in A.4}
{\bf 3}\cdot{\bf 3}\cdot{\bf 3}.
\end{equation}
When $a+b<0$, matter field $\overline{\bf 3}$ on the bulk $E$ and two matter fields $\overline{\bf 3}$ localized along the matter curve $\Sigma_0$ generate the following Yukawa coupling: 
\begin{equation}
\label{Yukawa 2 in A.4}
\overline{\bf 3}\cdot\overline{\bf 3}\cdot\overline{\bf 3}.
\end{equation}
The results are displayed in Table \ref{matter ci}.
\par We saw in Section \ref{ssec 3.5} that F-theory model on the Calabi-Yau complete intersection (\ref{eq type 3 in A.1}) has a discrete $\Z_2$ symmetry; therefore massless matter fields are charged under a discrete $\Z_2$ symmetry, and this has to be reflected in the structure of Yukawa couplings \cite{GGK}. Yukawa couplings (\ref{Yukawa 1 in A.4}) and (\ref{Yukawa 2 in A.4}) are indeed invariant under the action of $\Z_2$.

\begingroup
\renewcommand{\arraystretch}{1.5}
\begin{table}[htb]
\begin{flushleft}
  \begin{tabular}{|c|c|c|c|c|c|c|} \hline
Gauge Group & $a+b$ & Matters on $E$ & \# Gen. on $E$ & Matters on $\Sigma_0$ & \# Gen. on $\Sigma_0$ & Yukawa \\ \hline
$SU(4)$ & $>0$ & {\bf 3} & $8(a+b)$ & {\bf 3} & $2(a+b)$ & ${\bf 3}\cdot{\bf 3}\cdot{\bf 3}$ \\ \cline{2-7}
 & $<0$ & $\overline{\bf 3}$ & $-8(a+b)$ & $\overline{\bf 3}$ & $-2(a+b)$ & $\overline{\bf 3}\cdot\overline{\bf 3}\cdot\overline{\bf 3}$ \\ \hline
\end{tabular}
\caption{Potential matter spectra in F-theory compactification on (2,1,1,1) and (2,1,1,1) complete intersection (\ref{eq type 3 in A.1}) in $\P^3\times \P^1\times \P^1 \times \P^1$.}
\label{matter ci}
\end{flushleft}
\end{table}  
\endgroup


\begin{thebibliography}{99}
\bibitem{Vaf}C.~Vafa, ``Evidence for F-theory'', {\it Nucl. Phys.} {\bf B 469} (1996) 403 [arXiv:hep-th/9602022].
\bibitem{MV1}D.~R.~Morrison and C.~Vafa, ``Compactifications of F-theory on Calabi-Yau threefolds. 1'', {\it Nucl. Phys.} {\bf B 473} (1996) 74 [arXiv:hep-th/9602114].
\bibitem{MV2}D.~R.~Morrison and C.~Vafa, ``Compactifications of F-theory on Calabi-Yau threefolds. 2'', {\it Nucl. Phys.} {\bf B 476} (1996) 437 [arXiv:hep-th/9603161].

\bibitem{GW}T.~W.~Grimm and T.~Weigand, ``On Abelian Gauge Symmetries and Proton Decay in Global F-theory GUTs'', {\it Phys. Rev.} {\bf D82} (2010) 086009 [arXiv:1006.0226 [hep-th]].
\bibitem{MP}D.~R.~Morrison and D.~S.~Park, ``F-Theory and the Mordell-Weil Group of Elliptically-Fibered Calabi-Yau Threefolds'', {\it JHEP} {\bf 10} (2012) 128 [arXiv:1208.2695 [hep-th]].
\bibitem{CGK}M.~Cveti\v c, T.~W.~Grimm and D.~Klevers, ``Anomaly Cancellation And Abelian Gauge Symmetries In F-theory'', {\it JHEP} {\bf 02} (2013) 101 [arXiv:1210.6034 [hep-th]]. 
\bibitem{MPW}C.~Mayrhofer, E.~Palti and T.~Weigand, ``U(1) symmetries in F-theory GUTs with multiple sections'', {\it JHEP} {\bf 03} (2013) 098 [arXiv:1211.6742 [hep-th]].
\bibitem{BGK}V.~Braun, T.~W.~Grimm and J.~Keitel, ``New Global F-theory GUTs with U(1) symmetries'', {\it JHEP} {\bf 09} (2013) 154 [arXiv:1302.1854 [hep-th]].
\bibitem{BMPWsection}J.~Borchmann, C.~Mayrhofer, E.~Palti and T.~Weigand, ``Elliptic fibrations for $SU(5)\times U(1)\times U(1)$ F-theory vacua'', {\it Phys. Rev.} {\bf D88} (2013) no.4 046005 [arXiv:1303.5054 [hep-th]].
\bibitem{CKP}M.~Cveti\v c, D.~Klevers and H.~Piragua, ``F-Theory Compactifications with Multiple U(1)-Factors: Constructing Elliptic Fibrations with Rational Sections'', {\it JHEP} {\bf 06} (2013) 067 [arXiv:1303.6970 [hep-th]].
\bibitem{BGKgauge}V.~Braun, T.~W.~Grimm and J.~Keitel, ``	
Geometric Engineering in Toric F-Theory and GUTs with U(1) Gauge Factors'', {\it JHEP} {\bf 12} (2013) 069 [arXiv:1306.0577 [hep-th]]. 
\bibitem{CGKP}M.~Cveti\v c, A.~Grassi, D.~Klevers and H.~Piragua, ``Chiral Four-Dimensional F-Theory Compactifications With SU(5) and Multiple U(1)-Factors,'' {\it JHEP} {\bf 1404} (2014) 010 [arXiv:1306.3987 [hep-th]].
\bibitem{BMPWSU(5)}J.~Borchmann, C.~Mayrhofer, E.~Palti and T.~Weigand, ``SU(5) Tops with Multiple U(1)s in F-theory'', {\it Nucl.Phys.} {\bf B882} (2014) 1--69 [arXiv:1307.2902 [hep-th]]. 
\bibitem{CKPaddendum}M.~Cveti\v c, D.~Klevers and H.~Piragua, ``F-Theory Compactifications with Multiple U(1)-Factors: Addendum,'' {\it JHEP} {\bf 1312} (2013) 056 [arXiv:1307.6425 [hep-th]].
\bibitem{CKPS}M.~Cveti\v c, D.~Klevers, H.~Piragua and P.~Song, ``Elliptic fibrations with rank three Mordell-Weil group: F-theory with U(1) x U(1) x U(1) gauge symmetry,'' {\it JHEP} {\bf 1403} (2014) 021 [arXiv:1310.0463 [hep-th]].
\bibitem{BMPW}M.~Bies, C.~Mayrhofer, C.~Pehle and T.~Weigand, ``Chow groups, Deligne cohomology and massless matter in F-theory'', [arXiv:1402.5144 [hep-th]].
\bibitem{GK}T.~W.~Grimm and A.~Kapfer, ``Anomaly Cancelation in Field Theory and F-theory on a Circle'', {\it JHEP} {\bf 05} (2016) 102 [arXiv:1502.05398 [hep-th]]. 
\bibitem{LSW}C.~Lawrie, S.~Sch\"afer-Nameki and J.-M.~Wong, ``F-theory and All Things Rational: Surveying U(1) Symmetries with Rational Sections'', {\it JHEP} {\bf 09} (2015) 144 [arXiv:1504.05593 [hep-th]].
\bibitem{CKPT}M.~Cveti\v c, D.~Klevers, H.~Piragua and W.~Taylor, ``General U(1)$\times$U(1) F-theory compactifications and beyond: geometry of unHiggsings and novel matter structure,'' {\it JHEP} {\bf 1511} (2015) 204 [arXiv:1507.05954 [hep-th]].
\bibitem{CGKPS}M.~Cveti\v c, A.~Grassi, D.~Klevers, M.~Poretschkin and P.~Song, ``Origin of Abelian Gauge Symmetries in Heterotic/F-theory Duality,'' {\it JHEP} {\bf 1604} (2016) 041 [arXiv:1511.08208 [hep-th]].
\bibitem{MP2}D.~R.~Morrison and D.~S.~Park, ``Tall sections from non-minimal transformations'', {\it JHEP} {\bf 10} (2016) 033 [arXiv:1606.07444 [hep-th]]. 

\bibitem{BM}V.~Braun and D.~R.~Morrison, ``F-theory on Genus-One Fibrations'', {\it JHEP} {\bf 08} (2014) 132 [arXiv:1401.7844 [hep-th]].
\bibitem{MTsection}D.~R.~Morrison and W.~Taylor, ``Sections, multisections, and $U(1)$ fields in F-theory'', {\it J. Singularities} {\bf 15} (2016) 126--149 [arXiv:1404.1527 [hep-th]].
\bibitem{BEFNQ}P.~Berglund, J.~Ellis, A.~E.~Faraggi, D.~V.~Nanopoulos and Z.~Qiu, ``Elevating the free fermion $Z_2\times Z_2$ orbifold model to a compactification of F-theory'', {\it Int. Jour. of Mod. Phys.} {\bf A 15} (2000) 1345--1362 [arXiv:hep-th/9812141].
\bibitem{BDHKMMS}J.~de~Boer, R.~Dijkgraaf, K.~Hori, A.~Keurentjes, J.~Morgan, D.~R.~Morrison and S.~Sethi, ``Triples, fluxes, and strings'', {\it Adv. Theor. Math. Phys.} {\bf 4} (2002) 995--1186 [arXiv: hep-th/0103170].

\bibitem{AGGK}L.~B.~Anderson, I.~Garcia-Etxebarria, T.~W.~Grimm and J.~Keitel, ``Physics of F-theory compactifications without section'', {\it JHEP} {\bf 12} (2014) 156 [arXiv:1406.5180 [hep-th]].
\bibitem{KMOPR}D.~Klevers, D.~K.~Mayorga Pena, P.-K.~Oehlmann, H.~Piragua and J.~Reuter, ``F-Theory on all Toric Hypersurface Fibrations and its Higgs Branches'', {\it JHEP} {\bf 01} (2015) 142 [arXiv:1408.4808 [hep-th]].
\bibitem{GGK}I.~Garcia-Etxebarria, T.~W.~Grimm and J.~Keitel, ``Yukawas and discrete symmetries in F-theory compactifications without section'', {\it JHEP} {\bf 11} (2014) 125 [arXiv:1408.6448 [hep-th]].
\bibitem{MPTW}C.~Mayrhofer, E.~Palti, O.~Till and T.~Weigand, ``Discrete Gauge Symmetries by Higgsing in four-dimensional F-Theory Compactifications'', {\it JHEP} {\bf 12} (2014) 068 [arXiv:1408.6831 [hep-th]].
\bibitem{MPTW2}C.~Mayrhofer, E.~Palti, O.~Till and T.~Weigand, ``On Discrete Symmetries and Torsion Homology in F-Theory'', {\it JHEP} {\bf 06} (2015) 029 [arXiv:1410.7814 [hep-th]].
\bibitem{CDKPP}M.~Cveti\v c, R.~Donagi, D.~Klevers, H.~Piragua and M.~Poretschkin, ``F-theory vacua with $\mathbb Z_3$ gauge symmetry'', {\it Nucl. Phys.} {\bf B898} (2015) 736--750 [arXiv:1502.06953 [hep-th]].
\bibitem{LMTW}L.~Lin, C.~Mayrhofer, O.~Till and T.~Weigand, ``Fluxes in F-theory Compactifications on Genus-One Fibrations'', {\it JHEP} {\bf 01} (2016) 098 [arXiv:1508.00162 [hep-th]].
\bibitem{GKK}T.~W.~Grimm, A.~Kapfer and D.~Klevers, ``The Arithmetic of Elliptic Fibrations in Gauge Theories on a Circle'', {\it JHEP} {\bf 06} (2016) 112 [arXiv:1510.04281 [hep-th]].
\bibitem{K}Y.~Kimura, ``Gauge Groups and Matter Fields on Some Models of F-theory without Section'', {\it JHEP} {\bf 03} (2016) 042 [arXiv:1511.06912 [hep-th]].
\bibitem{K2}Y.~Kimura, ``Gauge groups and matter fields for F-theory without Section on Double Covers of $\P^1\times\P^1$ and Complete Intersections in $\P^1\times\P^3$'', [arXiv:1603.03212 [hep-th]].
\bibitem{KCY4}Y.~Kimura, ``F-theory Models without Section on Calabi-Yau 4-folds'', [arXiv:1607.02978 [hep-th]]. 

\bibitem{KNPRR}T.~Kobayashi, H.~P.~Nilles, F.~Ploger, S.~Raby and M.~Ratz, ``Stringy origin of non-Abelian discrete flavor symmetries,'' {\it Nucl.Phys.} {\bf B768} (2007) 135 [arXiv: hep-ph/0611020].
\bibitem{ACKO} H.~Abe, K.-S.~Choi, T.~Kobayashi and H.~Ohki, ``Non-Abelian Discrete Flavor Symmetries from Magnetized/Intersecting Brane Models,'' {\it Nucl.Phys.} {\bf B820} (2009) 317 [arXiv:0904.2631 [hep-ph]].
\bibitem{BS}T.~Banks and N.~Seiberg, ``Symmetries and Strings in Field Theory and Gravity'', {\it Phys. Rev.} {\bf D83} (2011) 084019 [arXiv:1011.5120 [hep-th]].
\bibitem{HSsums}S.~Hellerman and E.~Sharpe, ``Sums over topological sectors and quantization of Fayet-Iliopoulos parameters'', {\it Adv.Theor.Math.Phys.} {\bf 15} (2011) 1141--1199 [arXiv:1012.5999 [hep-th]].
\bibitem{CIM}P.~G.~Camara, L.~E.~Ibanez and F.~Marchesano, ``RR photons'', {\it JHEP} {\bf 09} (2011) 110 [arXiv:1106.0060 [hep-th]].
\bibitem{BISU}M.~Berasaluce-Gonzalez, L.~E.~Ibanez, P.~Soler and A.~M.~Uranga, ``Discrete gauge symmetries in D-brane models'', {\it JHEP} {\bf 12} (2011) 113 [arXiv:1106.4169 [hep-th]].
\bibitem{ISU}L.~E.~Ibanez, A.~N.~Schellekens and A.~M.~Uranga, ``Discrete Gauge Symmetries in Discrete MSSM-like Orientifolds'', {\it Nucl.Phys.} {\bf B865} (2012) 509--540 [arXiv:1205.5364 [hep-th]].
\bibitem{BCMRU}M.~Berasaluce-Gonzalez, P.~G.~Camara, F.~Marchesano, D.~Regalado and A.~M.~Uranga, ``Non-Abelian discrete gauge symmetries in 4d string models'', {\it JHEP} {\bf 09} (2012) 059 [arXiv:1206.2383 [hep-th]].
\bibitem{BCMU}M.~Berasaluce-Gonzalez, P.~G.~Camara, F.~Marchesano and A.~M.~Uranga, ``Zp charged branes in flux compactifications'', {\it JHEP} {\bf 04} (2013) 138 [arXiv:1211.5317 [hep-th]]. 
\bibitem{MRV}F.~Marchesano, D.~Regalado and L.~Vazquez-Mercado, ``Discrete flavor symmetries in D-brane models'', {\it JHEP} {\bf 09} (2013) 028 [arXiv:1306.1284 [hep-th]].
\bibitem{HS}G.~Honecker and W.~Staessens, ``To Tilt or Not To Tilt: Discrete Gauge Symmetries in Global Intersecting D-Brane Models'', {\it JHEP} {\bf 10} (2013) 146 [arXiv:1303.4415 [hep-th]].
\bibitem{BRU}M.~Berasaluce-Gonzalez, G.~Ramirez and A.~M.~Uranga, ``Antisymmetric tensor $Z_p$ gauge symmetries in field theory and string theory'', {\it JHEP} {\bf 01} (2014) 059 [arXiv:1310.5582 [hep-th]].
\bibitem{KKLM}A.~Karozas, S.~F.~King, G.~K.~Leontaris and A.~Meadowcroft, ``Discrete Family Symmetry from F-Theory GUTs'', {\it JHEP} {\bf 09} (2014) 107 [arXiv:1406.6290 [hep-ph]]. 
\bibitem{BGKintfiber}V.~Braun, T.~W.~Grimm and J.~Keitel, ``Complete Intersection Fibers in F-Theory'', {\it JHEP} {\bf 03} (2015) 125 [arXiv:1411.2615 [hep-th]]. 
\bibitem{HS2}G.~Honecker and W.~Staessens, ``Discrete Abelian gauge symmetries and axions'', {\it J.Phys.Conf.Ser.} {\bf 631} (2015) no.1 [arXiv:1502.00985 [hep-th]].
\bibitem{GPR}T.~W.~Grimm, T.~G.~Pugh and D.~Regalado, ``Non-Abelian discrete gauge symmetries in F-theory'', {\it JHEP} {\bf 02} (2016) 066 [arXiv:1504.06272 [hep-th]]. 
\bibitem{CGP}M.~Cveti\v c, A.~Grassi and M.~Poretschkin, ``Discrete Symmetries in Heterotic/F-theory Duality and Mirror Symmetry,'' [arXiv:1607.03176 [hep-th]]. 

\bibitem{Isk}V.~A.~Iskovskih, ``FANO 3-FOLDS. I'', {\it Math. USSR Izv.} {\bf 11} (1977) 485. 
\bibitem{Fujita1}T.~Fujita, ``On the structure of polarized manifolds with total deficiency one, I'', {\it J. Math. Soc. Japan} {\bf 32} (1980) 709--725.
\bibitem{Fujita2}T.~Fujita, ``On the structure of polarized manifolds with total deficiency one, II'', {\it J. Math. Soc. Japan} {\bf 33} (1981) 415--434.

\bibitem{Kod1}K.~Kodaira, ``On compact analytic surfaces II'', {\it Ann. of Math.} {\bf 77} (1963), 563--626.
\bibitem{Kod2}K.~Kodaira, ``On compact analytic surfaces III'', {\it Ann. of Math.} {\bf 78} (1963), 1--40.
\bibitem{Ner}A.~N\'eron, ``Mod\`eles minimaux des vari\'et\'es ab\'eliennes sur les corps locaux et globaux'', {\it Publ. math. de l'IH{\' E}S} {\bf 21} (1964), 5--125.

\bibitem{AM}P.~S.~Aspinwall and D.~R.~Morrison, ``Nonsimply connected gauge groups and rational points on elliptic curves'', {\it JHEP} {\bf 9807} (1998) 012 [arXiv: hep-th/9805206].
\bibitem{MMTW}C.~Mayrhofer, D.~R.~Morrison, O.~Till and T.~Weigand, ``Mordell-Weil Torsion and the Global Structure of Gauge Groups in F-theory'', {\it JHEP} {\bf 10} (2014) 16 [arXiv:1405.3656 [hep-th]].

\bibitem{GVW}S.~Gukov, C.~Vafa and E.~Witten, ``CFT's from Calabi-Yau four folds'', {\it Nucl. Phys.} {\bf B584} (2000) 69--108 [arXiv: hep-th/9906070].
\bibitem{BCV}A.~P.~Braun, A.~Collinucci and R.~Valandro, ``G-flux in F-theory and algebraic cycles'', {\it Nucl. Phys.} {\bf B 856} (2012) 129 [arXiv:1107.5337 [hep-th]].
\bibitem{MS}J.~Marsano and S.~Sch\"afer-Nameki, ``Yukawas, G-flux, and Spectral Covers from Resolved Calabi-Yau's'', {\it JHEP} {\bf 11} (2011) 098 [arXiv:1108.1794 [hep-th]].  
\bibitem{KMW}S.~Krause, C.~Mayrhofer and T.~Weigand, ``$G_4$ flux, chiral matter and singularity resolution in F-theory compactifications'', {\it Nucl. Phys.} {\bf B 858} (2012) 1--47 [arXiv:1109.3454 [hep-th]].
\bibitem{KMW2}S.~Krause, C.~Mayrhofer and T.~Weigand, ``Gauge Fluxes in F-theory and Type IIB Orientifolds'', {\it JHEP} {\bf 08} (2012) 119 [arXiv:1202.3138 [hep-th]]. 
\bibitem{BKL}N.~Cabo Bizet, A.~Klemm and D.~Vieira Lopes, ``	
Landscaping with fluxes and the E8 Yukawa Point in F-theory'', [arXiv:1404.7645 [hep-th]]. 
\bibitem{CKPOR}M.~Cveti\v c, D.~Klevers, D.~K.~M.~Pe$\tilde{\rm n}$a, P.-K.~Oehlmann and J.~Reuter, ``Three-Family Particle Physics Models from Global F-theory Compactifications'', {\it JHEP} {\bf 08} (2015) 087 [arXiv:1503.02068 [hep-th]].
\bibitem{SNW}S.~Sch\"afer-Nameki and T.~Weigand, ``F-theory and 2d (0,2) Theories'', {\it JHEP} {\bf 05} (2016) 059 [arXiv:1601.02015 [hep-th]].
\bibitem{LW}L.~Lin and T.~Weigand, ``$G_4$-Flux and Standard Model Vacua in F-theory'', {\it Nucl.Phys.} {\bf B913} (2016) 209--247 [arXiv:1604.04292 [hep-th]].

\bibitem{BB}K.~Becker and M.~Becker, ``M theory on eight manifolds'', {\it Nucl. Phys.} {\bf B477} (1996) 155--167 [arXiv: hep-th/9605053].
\bibitem{SVW}S.~Sethi, C.~Vafa and E.~Witten, ``Constraints on low dimensional string compactifications'', {\it Nucl. Phys.} {\bf B 480} (1996) 213--224, [arXiv: hep-th/9606122].
\bibitem{W}E.~Witten, ``On flux quantization in M theory and the effective action'', {\it J. Geom. Phys.} {\bf 22} (1997) 1--13 [arXiv: hep-th/9609122].
\bibitem{DRS}K.~Dasgupta, G.~Rajesh and S.~Sethi, ``M theory, orientifolds and G-flux'', {\it JHEP} {\bf 08} (1999) 023 [arXiv: hep-th/9908088]. 

\bibitem{SchShio}M.~Sch\"{u}tt and T.~Shioda, ``Elliptic Surfaces'', in {\it Algebraic Geometry in East Asia (Seoul 2008)}, {\it Advanced Studies in Pure Mathematics} {\bf 60} (2010) 51--160 [arXiv:0907.0298 [math.AG]].

\bibitem{BIKMSV}M.~Bershadsky, K.~A.~Intriligator, S.~Kachru, D.~R.~Morrison, V.~Sadov and C.~Vafa, ``Geometric singularities and enhanced gauge symmetries'', {\it Nucl. Phys.} {\bf B 481} (1996) 215 [arXiv:hep-th/9605200].
\bibitem{KV}S.~H.~Katz and C.~Vafa, ``Matter from geometry'', {\it Nucl. Phys.} {\bf B 497} (1997) 146 [arXiv:hep-th/9606086].
\bibitem{BHV1}C.~Beasley, J.~J.~Heckman and C.~Vafa, ``GUTs and Exceptional Branes in  F-theory -I'', {\it JHEP} {\bf 01} (2009) 058 [arXiv:0802.3391 [hep-th]].
\bibitem{Sla}R.~Slansky, ``Group theory for unified model building'', {\it Physics Reports} {\bf 79} (1981) 1--128. 


\end{thebibliography}
\end{document}